\documentclass[journal,12pt,onecolumn,draftclsnofoot,]{IEEEtran}

\usepackage{float}
\usepackage{graphicx}
\usepackage{subfigure}
\usepackage{array}
\usepackage{longtable}
\usepackage{amssymb}
\usepackage{amsmath}
\usepackage{amsthm}
\usepackage{amsfonts}
\usepackage{mathrsfs}
\usepackage{fancyhdr}
\usepackage{cite}
\usepackage{gensymb}
\usepackage{enumerate}
\usepackage{setspace}
\usepackage{color}
\usepackage{algorithm}
\usepackage{algorithmic}
\usepackage{booktabs}
\usepackage{bm}
\usepackage{multirow}

\usepackage[colorlinks,
            linkcolor=blue,
            anchorcolor=blue,
            citecolor=blue
            ]{hyperref}

\newtheorem{proposition}{Proposition}

\begin{document}

\hyphenpenalty=5000 \tolerance=1000

\title{\LARGE Probabilistic Tile Visibility-Based Server-Side Rate Adaptation for Adaptive 360-Degree Video Streaming}

\author{{\small Junni~Zou,}~\IEEEmembership{\small Member,~IEEE,}  {\small Chenglin~Li,}~\IEEEmembership{\small Member,~IEEE,} {\small Chengming~Liu,} {\small Qin~Yang,} {\small Hongkai~Xiong,}~\IEEEmembership{\small Senior Member,~IEEE,} {\small and Eckehard~Steinbach,} \IEEEmembership{\small Fellow,~IEEE}
\thanks{


\footnotesize{J.~Zou, C.~Li, Q.~Yang and H.~Xiong are with the School of Electronic Information and Electrical Engineering, Shanghai Jiao Tong University, Shanghai 200240, China (e-mail: zou-jn@cs.sjtu.edu.cn, lcl1985@sjtu.edu.cn, yangqin@sjtu.edu.cn, xionghongkai@sjtu.edu.cn).}

\footnotesize{C.~Liu is with the Dept. of Communication Engineering, Shanghai University, Shanghai 200240, China (e-mail: cmliu@shu.edu.cn).}

\footnotesize{E.~Steinbach is with the Chair of Media Technology (LMT), Technical University of Munich (TUM), Munich 80333, Germany (e-mail: eckehard.steinbach@tum.de).}
}
}

\maketitle \vspace*{-1.2cm}

\begin{abstract}
In this paper, we study the server-side rate adaptation problem for streaming tile-based adaptive 360-degree videos to multiple users who are competing for transmission resources at the network bottleneck. Specifically, we develop a convolutional neural network (CNN)-based viewpoint prediction model to capture the nonlinear relationship between the future and historical viewpoints. A Laplace distribution model is utilized to characterize the probability distribution of the prediction error. Given the predicted viewpoint, we then map the viewport in the spherical space into its corresponding planar projection in the 2-D plane, and further derive the visibility probability of each tile based on the planar projection and the prediction error probability. According to the visibility probability, tiles are classified as viewport, marginal and invisible tiles. The server-side tile rate allocation problem for multiple users is then formulated as a non-linear discrete optimization problem to minimize the overall received video distortion of all users and the quality difference between the viewport and marginal tiles of each user, subject to the transmission capacity constraints and users' specific viewport requirements. We develop a steepest descent algorithm to solve this non-linear discrete optimization problem, by initializing the feasible starting point in accordance with the optimal solution of its continuous relaxation. Extensive experimental results show that the proposed algorithm can achieve a near-optimal solution, and outperforms the existing rate adaptation schemes for tile-based adaptive 360-video streaming.
\end{abstract}



\section{Introduction}
\label{sec:intro}

In recent years, watching 360-degree videos with head-mounted displays (HMDs) has become a popular virtual reality (VR) application. Compared to traditional videos, 360-degree videos provide users with a panoramic scene captured by an omnidirectional camera. When watching 360-degree videos, the user is able to obtain an immersive experience by freely adjusting her/his viewing orientation. Due to its huge file size and ultra high resolution, the delivery of a 360-degree video may consume up to six times the transmission rate of a traditional video \cite{Bao2017}. Current network infrastructures, especially mobile networks, can hardly support the full delivery of the entire 360-degree video to users. An intuitive solution is to reduce the encoding and thus the transmission bitrate of 360-degree videos, which, however, would inevitably degrade the visual quality. Therefore, how to effectively save transmission rate while preserving the quality of experience (QoE) for the users is a challenging problem in the field of 360-degree video streaming.

Constrained by the field of view (FoV) of the HMD (e.g., 90-degree vertically and 110-degree horizontally), the user at any time can only view a small portion of the full 360-degree scene, which is called the \textit{viewport} of the user. Thus, streaming only the viewport of the user at a high quality provides an effective rate-saving approach. In practice, such a transmission rate adaptability can be realized by integrating the concept of tiling \cite{Girod2009} with HTTP adaptive streaming (HAS) techniques. In tile-based HAS, a 360-degree video is divided into several spatial tiles and each spatial tile is further divided into several temporal segments, then each tile in each segment can be encoded and delivered at different quality levels, in adaptation to the user's interest and rate constraint.

Tile-based HAS requires to predict the user's viewport and prefetch tiles within the predicted viewport in advance. For viewport prediction, a general approach is to extrapolate the current viewport into the future by using historical viewports. Since a variety of factors, such as preference, occupation, gender, and age, influence the viewport of interest, the relationship between the future and historical viewports can be characterized as nonlinear and long-term dependent. Existing prediction methods, e.g., based on linear regression (LR) and neural networks (NNs) \cite{Bao2017}, fail to capture these properties well, which may result in an undesirable prediction error.

If the viewport prediction can achieve $100\%$ accuracy, it is surely rate-efficient to stream the tiles inside the predicted viewport at a higher quality and other invisible tiles at a lower quality. However, absolutely accurate viewport prediction can hardly be reached, especially when the user is fast moving her/his head. In this case, some marginal tiles that compensate for the prediction error between the predicted viewport and the user's actual viewport need to be transmitted at a moderate quality for smooth playback on the user side. Intuitively, a larger margin will lead to a higher probability that the user's actual viewport can be fully covered, but with a cost of more transmission rate consumed. This paper focuses on the tile rate allocation for both the predicted viewport and the margin area based on the visibility probability of different tiles, so as to seek the tradeoff between the user QoE and transmission rate utilization.

For adaptive 360-degree video streaming, the rate adaptation schemes in the literature \cite{Laura2017, Qian2016, Ghosh2017, Corbillon2017, Kan2019} are mostly user-driven. Namely, users determine the best encoding bitrates of tiles to download from the server based on their transmission capacity estimation and buffer occupancy. Such a user-side rate adaptation, despite its popularity, has shown some disadvantages. For example, the user makes the rate adaptation decision merely based on local information, without considering the conditions of the server and network. When multiple users compete for the transmission capacity at the network bottleneck, their QoE, in terms of stability, fairness and efficiency, achieved by the user-side rate adaptation becomes sub-optimal \cite{Marai2018, Akhshabi2013, Li2014}. Furthermore, the current HAS is used over HTTP/1.1, such as dynamic adaptive streaming over HTTP (DASH) \cite{Stockhammer2011}, which follows a pull-based video retrieval. When DASH is used for tile-based adaptive 360-degree video streaming, the user needs to send an independent request for each video tile, leading to an undesirable overhead and latency. The newly published HTTP/2 protocol \cite{HTTP2} overcomes the above drawbacks of HTTP/1.1 by using a server-side push technique. It allows the server to push multiple content to a user instead of making individual request for each content, which therefore enables server-side rate adaptation. In practice, HTTP/2 is easy to be realized as it shares the same semantics and keeps backward compatibility with HTTP/1.1 \cite{Petrangeli2017, Xiao2018}.

In this paper, we study the server-side rate adaptation for multiple users who are competing for the server's transmission capacity as the network bottleneck, subject to their personal rate constraints and specific viewport requirements. Based on the viewpoint/viewport prediction for users, the mapping from the spherical viewport to its corresponding planar projection, and the corresponding visibility probability derivation of each tile for each user, the server then determines the encoding/transmission rate of each tile to be sent to the users, aiming at minimizing the overall expected video distortion perceived by all the users. The main contributions of this paper can be summarized as follows.

1) Existing works on viewport prediction usually predict the center point of the viewport, i.e., a kind of \textit{viewpoint} prediction. To improve the viewpoint prediction accuracy, in particular, to more accurately capture the nonlinear relationship between the future and past viewpoints, we develop a convolutional neural network (CNN)-based viewing angle prediction model, in which the pooling layers are dropped and more convolutional layers are added for stronger nonlinear fitting ability. Experimental results show that our model outperforms the previous works, especially for large-size prediction windows.

2) Given a predicted viewpoint/viewport on the spherical space, how to find its corresponding area in the 2-D projection plane with the set of covering tiles has still remained unexplored for 360-degree video streaming. In this paper, we theoretically analyze how to map the user's viewport on the sphere to the 2-D projection plane and determine the set of tiles within the 2-D projection plane that cover this viewport. To the best of our knowledge, such a mapping from the viewport on the sphere to the corresponding area in the 2-D projection plane has barely been studied in the literature.

3) Based on real head movement traces, we adopt a Laplace distribution model to characterize the probability of viewpoint prediction error, with which the viewpoint prediction error can be presented more accurately. We further derive the visibility probability of each tile, based on which the tiles are classified into viewport, marginal and invisible tiles to support tile-based rate adaptation.

4) We develop an optimal server-side rate adaptation framework, in which the tile rate allocation optimization problem among multiple users is formulated as a non-linear discrete optimization problem, aiming at maximizing the received video quality and navigation quality smoothness of multiple users. Thereafter, a steepest descent algorithm is developed to solve this non-linear discrete optimization problem, where the feasible starting point is determined by the optimal solution of its continuous relaxation.

The remainder of this paper is organized as follows. Section \ref{sec:relatewk} reviews the related work on adaptive 360-degree video streaming. Section \ref{sec:viewport} presents the design of the CNN-based viewpoint prediction model. In Section \ref{sec:probability}, we propose the mapping from the user's viewport on the sphere to the corresponding area in the 2-D projection plane, and derive a probabilistic tile visibility model. In Section \ref{sec:rate}, the optimal server-side rate adaptation problem is formulated and solved with the steepest descent algorithm. Simulation results are presented and discussed in Section \ref{sec:simulate}. Finally, Section \ref{sec:conslusion} concludes this paper.

\section{Related Work}
\label{sec:relatewk}

Rate adaptation for 360-degree video streaming involves three sequential procedures: viewpoint/viewport prediction, mapping from the predicted viewpoint/viewport to the planar tile set of interest, and rate allocation among the tiles. Most of the existing works separately studied the first and the third procedure, leaving the second procedure untouched.

For the viewpoint/viewport prediction, Qian \textit{et al.} \cite{Qian2016} proposed a naive prediction model that directly utilizes the current viewpoint of the user to represent her/his future viewpoint. Bao \textit{et al.} \cite{Bao2016} proposed a linear regression model and a neural network to fit the variation of the user's viewpoint. Azuma \textit{et al.} \cite{Azuma1995} characterized the user's head motion as position, velocity and acceleration, and proposed a predictor to derive the future head position. The authors in \cite{Aladagli2018} took content-related features into account, and predicted the viewpoint based on a saliency algorithm. A deep reinforcement learning based viewpoint prediction approach is proposed in \cite{Xu2018}, in which the reinforcement learning model is established to track the long-term head movement behaviors of humans.

The rate allocation schemes in the literature \cite{Laura2017, Qian2016, Ghosh2017, Corbillon2017, Kan2019} usually assume that the tile set of interest is given and directly consider rate adaptation for this tile set. For instance, Toni \textit{et al.} \cite{Laura2017} proposed a tile-based adaptive streaming strategy to determine the rate at which each tile is downloaded for maximizing the quality experienced, where the tile set required by the user is given. Given the viewport and bandwidth estimation, Ghosh \textit{et al.} \cite{Ghosh2017} formulated different QoE metrics and designed a streaming algorithm for 360-degree video streaming. The authors in \cite{Corbillon2017} presented a viewport-adaptive 360-degree video streaming system, in which the front face is encoded in full quality while the other faces are encoded in low quality. Zou \textit{et al.} \cite{Kan2019} proposed a deep reinforcement learning-based rate adaptation algorithm to maximize the user QoE by adapting the transmitted video quality to the time-varying network conditions, which also assumes tile set of interest known.

This work differs from the related literature in the following aspects. First, we study all these three procedures and present a complete framework including viewpoint prediction, tile set mapping and rate adaptation. Second, considering that convolutional neural network (CNN) can well capture the nonlinear relationship between the future and past viewpoints, we establish a CNN-based viewpoint prediction model to generate predicted data for the tile set mapping and rate adaptation procedure. It is worth mentioning that the proposed tile mapping and rate adaptation strategy can be compatible with any viewpoint prediction approach in the literature. Third, we seek the rate adaptation from the server-side for a general scenario, where multiple users simultaneously send their video requests to the server, resulting in a rate competition on the server side.

\section{CNN-based Viewpoint Prediction}
\label{sec:viewport}

\begin{figure}[!t]
    \centering
    \subfigure[]{\label{fig_a}
    \includegraphics[width=2.1in]{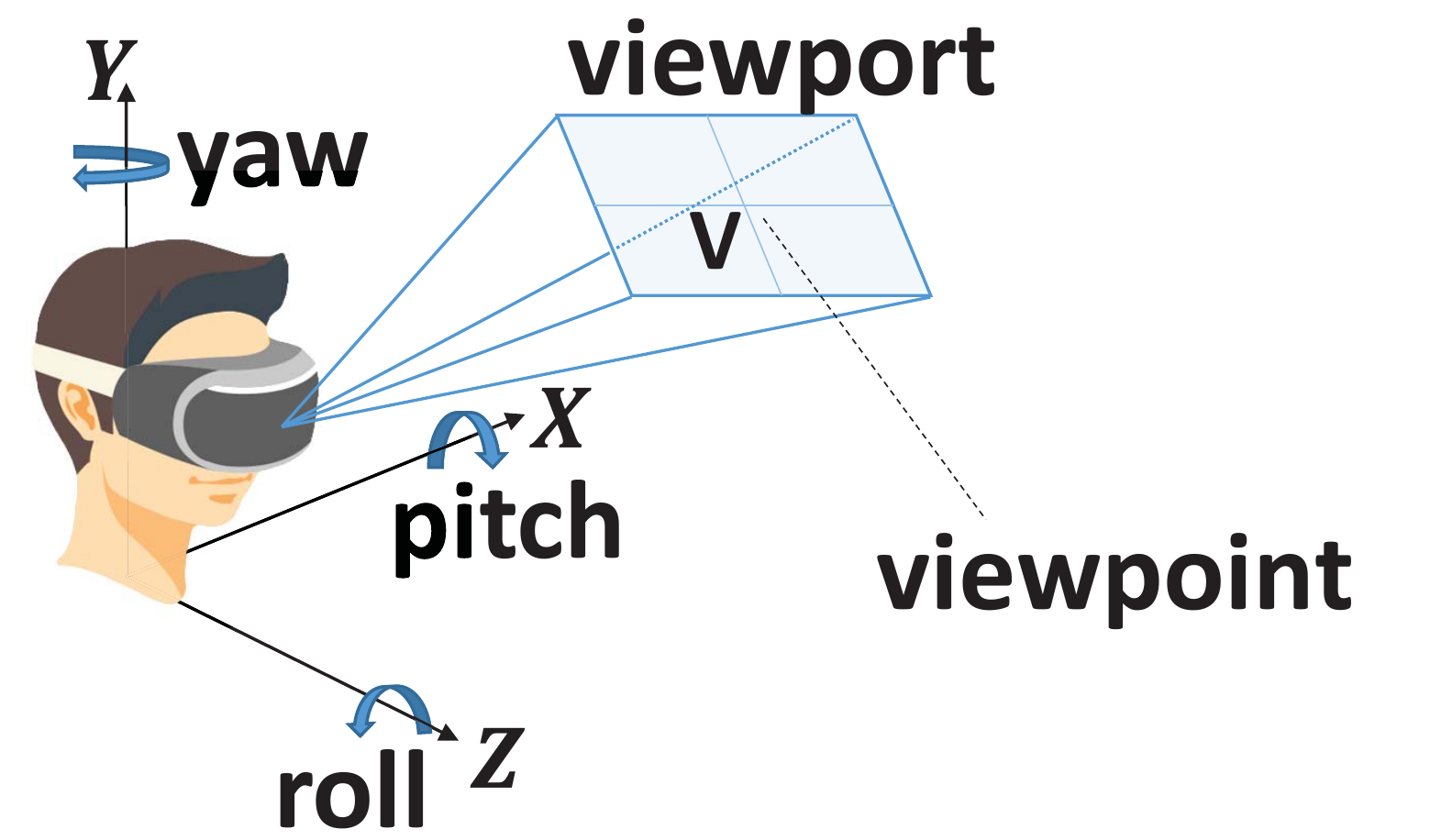}}
    \subfigure[]{\label{fig_b}
    \includegraphics[width=1.65in]{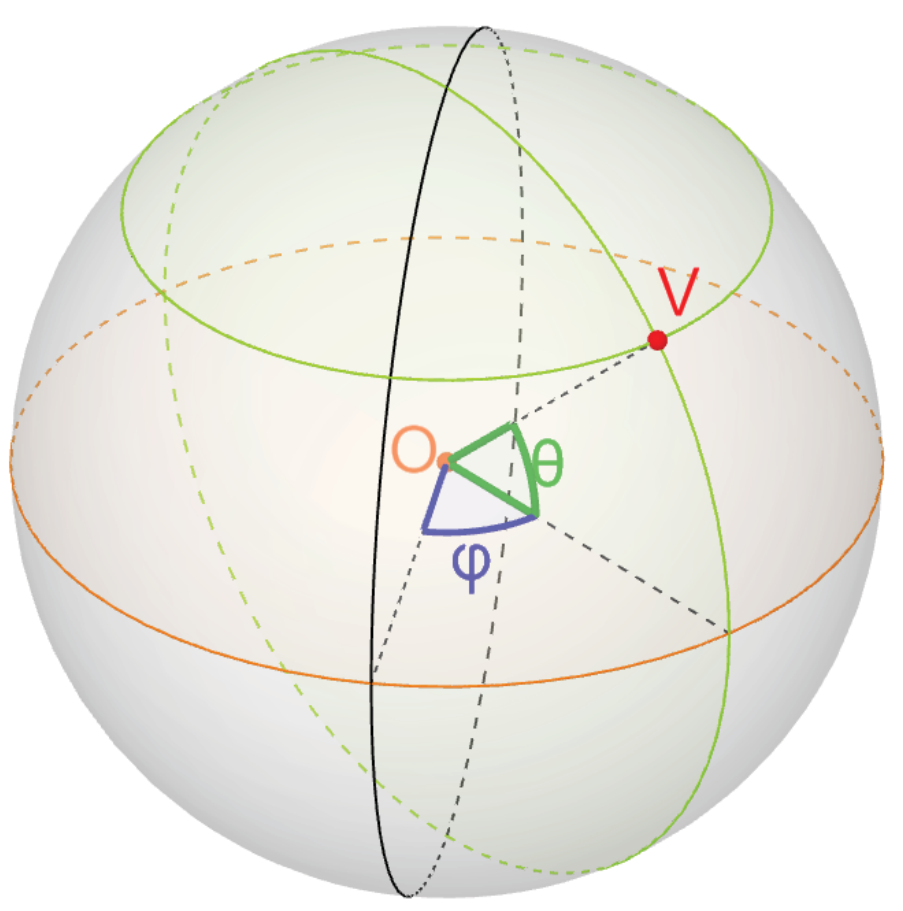}}
    \caption{(a) Head rotation angles; and (b) the position of viewpoint is represented by latitude $\theta$ and longitude $\varphi$ in spherical space with the rotations of pitch and yaw equal to the latitude and longitude, respectively.}
    \label{fig:rotation}\vspace*{-0.4cm}
    \end{figure}

As illustrated in Fig. \ref{fig:rotation}, when watching a 360-degree video, the user wearing an HMD is supposed to stand at the center point $O$ of the sphere, with her/his viewpoint (i.e., the center point $V$ of the viewport) represented by the Euler angles, pitch ($\theta$), yaw ($\varphi$), and roll ($\psi$), respectively, corresponding to the head rotation around the $X$, $Y$ and $Z$ axis. Knowing $\theta$ and $\varphi$, the user's viewpoint can be determined as shown in Fig. \ref{fig:rotation}(a). We define the initial head rotation as zero degree for pitch, yaw, and roll angles, then $\theta$ and $\psi$ rotate in a range of $[-90^{\circ}, 90^{\circ}]$, $\varphi$ rotates in a range of $[-180^{\circ}, 180^{\circ}]$.

%

Let $\mathcal{R}_t=(\theta_t,\varphi_t,\psi_t)$ represent the viewport of the user at time $t$, the goal of viewport prediction is to estimate the future viewport $\mathcal{R}_{t+t_w}$ for a given series of historical head rotations $\mathcal{R}_{t-t_s}$, ..., $\mathcal{R}_{t-1}$, $\mathcal{R}_{t}$, where $t_s$ is the time span of the considered historical views and $t_w$ is the size of the prediction window. The experimental results in \cite{Bao2016} show that, compared with the auto-correlations of these three angles, their cross-correlations are small and thus can be neglected. Therefore, we assume that the rotations in the three directions are independent of each other, which suggests that we can predict each angle independently by training three separate models. Further, it has been verified that the rotation in the direction of roll is negligible compared to the other two directions \cite{Bao2016}. For the sake of simplicity and without loss of generality, we assume that the roll angle stays at $0^{\circ}$ all the time when the user watches the video. Our task then becomes the estimation of the pitch and yaw angles, with the previous viewport prediction problem reducing to a viewpoint prediction problem as shown in Fig. \ref{fig:rotation}(a).

The design of a CNN-based viewing angle prediction model aims at building a CNN model that takes the current and past viewing angles (i.e., the features) as input and outputs the future viewing angles. In the following, we take the yaw angle $\varphi$ as an example to illustrate our model design. For the pitch angle, the CNN-based angle model can be developed in a similar way. We denote by $\varphi_{t}$ the yaw angle at time $t$, and $\pmb{\varphi}_{t-t_s:t} = (\varphi_{t-t_s},..., \varphi_{t-1},  \varphi_t)$ the yaw angles from time $t-t_s$ to time $t$ that are collected from the HMD sensor. The task of the CNN model becomes to predict $\varphi_{t+t_w}$ at some future point $t+t_w$ based on the values of $\pmb{\varphi}_{t-t_s:t}$.

According to our definition for the yaw angle, $-180^{\circ}$ and $179^{\circ}$ will just have a difference of $1^{\circ}$ instead of $359^{\circ}$. To address this issue, we take an angle transformation, and use $\mathbf{v}_t=(v_t^s,v_t^c)$ rather than $\varphi_t$ as the input. That is
\begin{align}
g(\varphi_t) &=(\sin(\varphi_t),\cos(\varphi_t)) \triangleq (v_t^s,v_t^c).
\end{align}
Before outputting the prediction result, we take the inverse transformation and obtain $\varphi_t$ from $\mathbf{v}_t$. Namely, $\varphi_t =\arctan(v_t^s / v_t^c)$.

\begin{figure}[!t]
 \centering
  \includegraphics[width=0.6\linewidth]{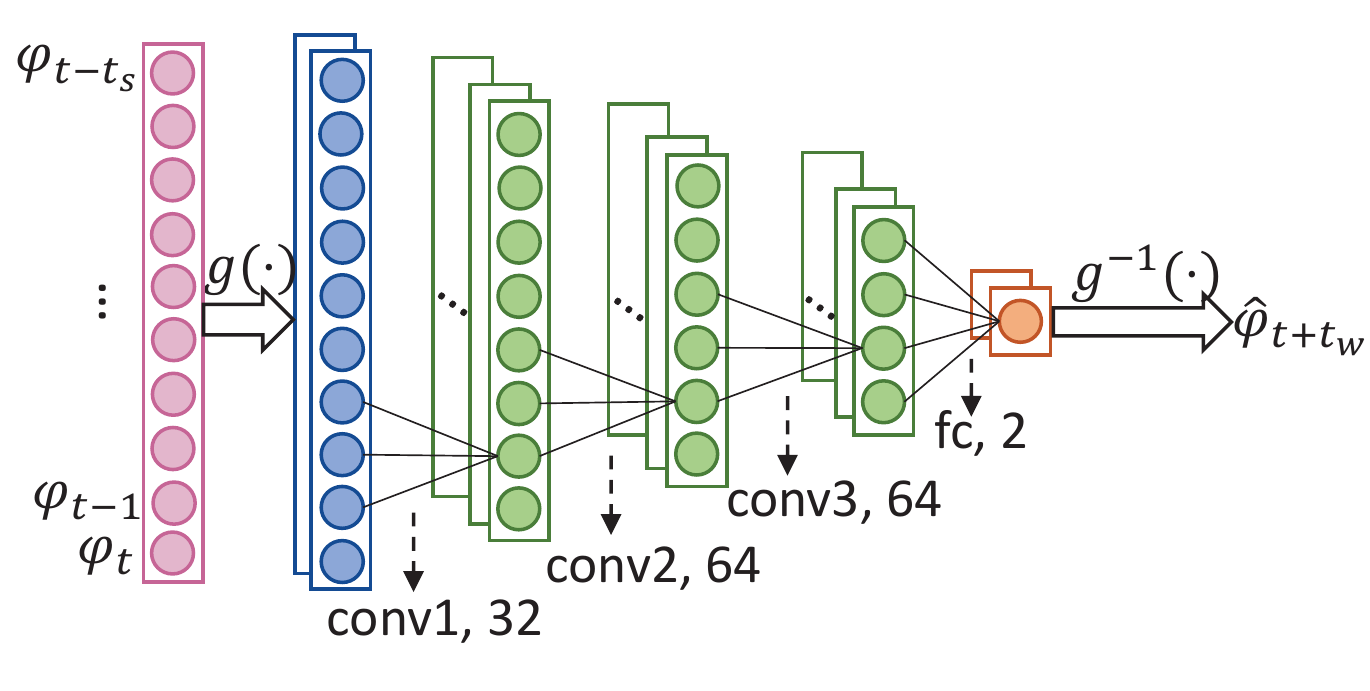}
  \caption{CNN-based viewing angle prediction model. }
  \label{fig:CNN}\vspace*{-0.4cm}
\end{figure}

Since the value of $\varphi_{t+t_w}$ exhibits strong nonlinear correlation with $\pmb{\varphi}_{t-t_s:t}$, we abandon the pooling layers to construct more convolutional layers so as to obtain stronger nonlinear fitting ability, as shown in Fig. \ref{fig:CNN}. Furthermore, we set the size of all kernels to 3 and the stride size to 1 without padding. Therefore, the depth of the network only depends on the input size. To find the optimal input size, we try the input size to be {5, 7, 10, 12, 15}, and the convolutional layers as {2, 3, 3, 5, 7}, respectively, and find that the highest prediction accuracy is achieved when the input size is set to 10.

Since the values of $v_{t}^s$ and $v_{t}^c$ range from $-1$ to $1$, we choose the hyperbolic tangent function $tanh()$ as the activation function of the fully connected layer, such that the output ranges from $-1$ to $1$. For all the convolutional layers, we set the activation function to be Rectified Linear Unit (ReLU), i.e., $f(x)=\max(0,x)$. In this work, we use mean squared error (MSE) as the loss function.

\section{Probabilistic Tile Visibility}
\label{sec:probability}
The angle prediction models provide us with the user's future viewpoint on the sphere. To perform tile rate allocation, what we need in the 2-D plane is the set of tiles corresponding to the user's viewport. In this section, we first present our method to determine the viewport tile region for a given spherical viewport. Based on the distribution of prediction error in the viewing angles, we further analyze the probability of tile visibility.

\subsection{From Spherical Viewport to Planar Viewport Tile Region}
\begin{figure}[!t]
 \centering
  \includegraphics[width=0.6\linewidth]{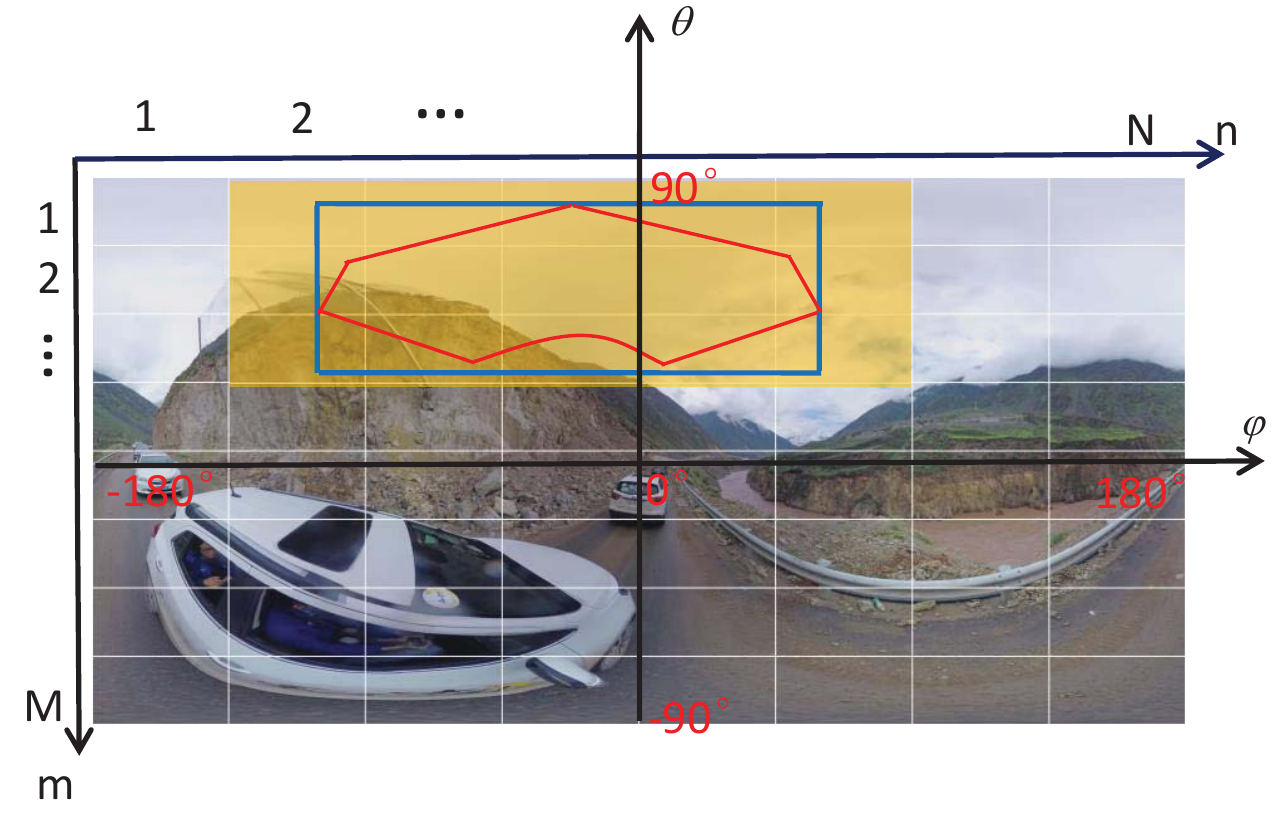}
  \caption{ Viewport and viewport tile region in the 2-D projection plane. }
  \label{fig:viewport}\vspace*{-0.2cm}
\end{figure}

Consider a 360-degree spherical video that is projected into a rectangular 2-D planar video by using the popular equirectangular projection (ERP) method. In order to unify the coordinate system, we introduce the latitude and longitude to uniformly represent the position of any point in both the spherical space and 2-D projection plane. Suppose that the user's viewpoint is $\mathcal{V}=(\theta,\varphi)$, namely, his viewpoint in the spherical space is located at $(\theta, \varphi)$, with a latitude $\theta$ and a longitude $\varphi$, as shown in Fig. \ref{fig:rotation}(b). In the 2-D plane, as seen in Fig. \ref{fig:viewport}, we use the horizontal lines to represent the latitudes of the spherical surface, and the vertical lines to denote the longitudes. Also, we let the longitudes of the eastern hemisphere to be positive, and those of the western hemisphere to be negative. Similarly, the latitudes of the northern hemisphere are set to be positive, and to be negative for the southern hemisphere. Following this mapping, the relationship between the viewport and the set of tiles within or overlapped with the viewport (referred to as \textit{viewport tile region}) can be depicted by Fig. \ref{fig:viewport}, where the region enclosed by the red line represents the projected user viewport, and the tiles with yellow color constitute the viewport tile region.

\begin{figure}[!t]
   \centering
\subfigure[ ]
{
  \begin{minipage}{0.415\linewidth}
    \includegraphics[width=\textwidth]{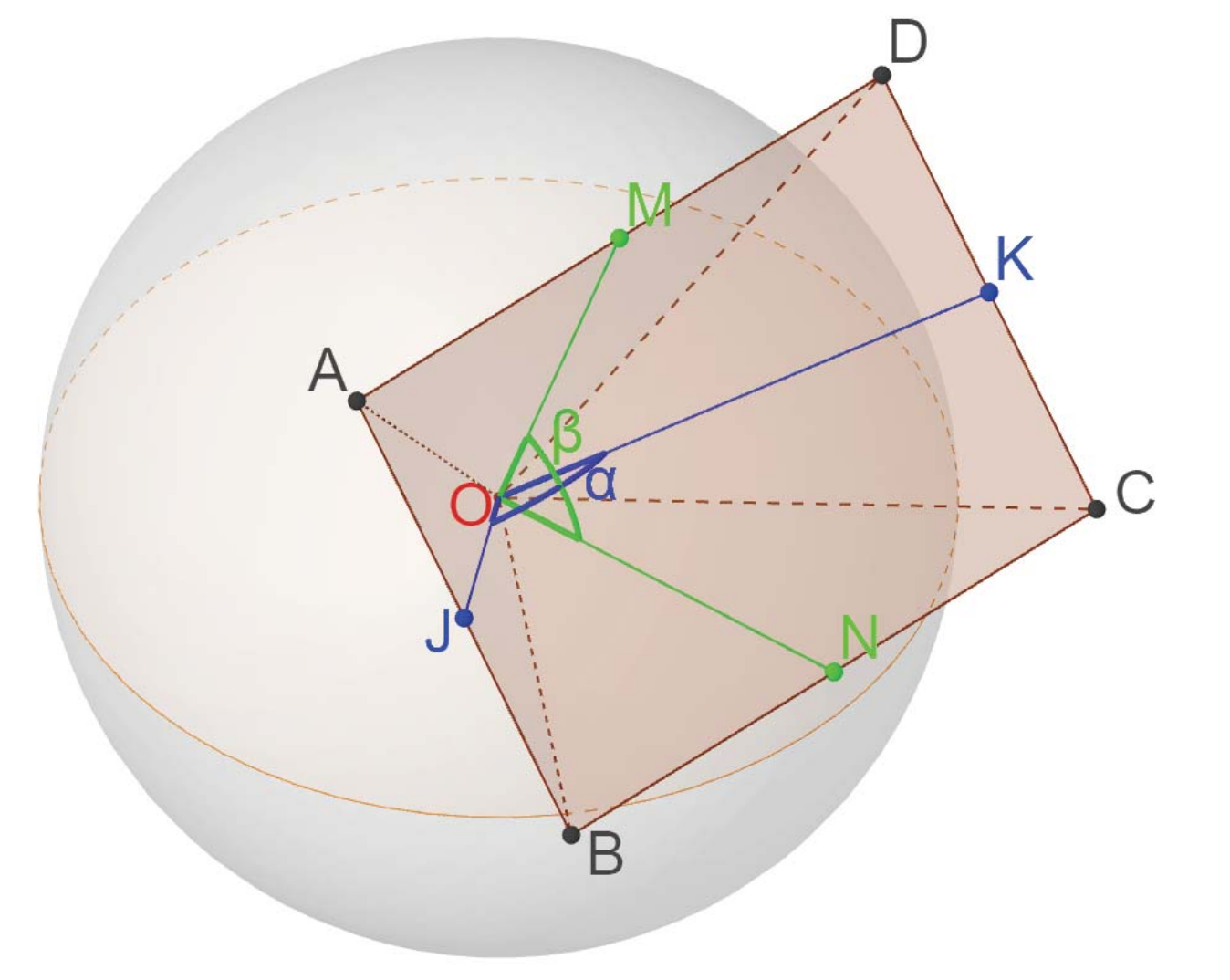}
  \end{minipage}
}
\subfigure[ ]
{
  \begin{minipage}{0.4\linewidth}
    \includegraphics[width=\textwidth]{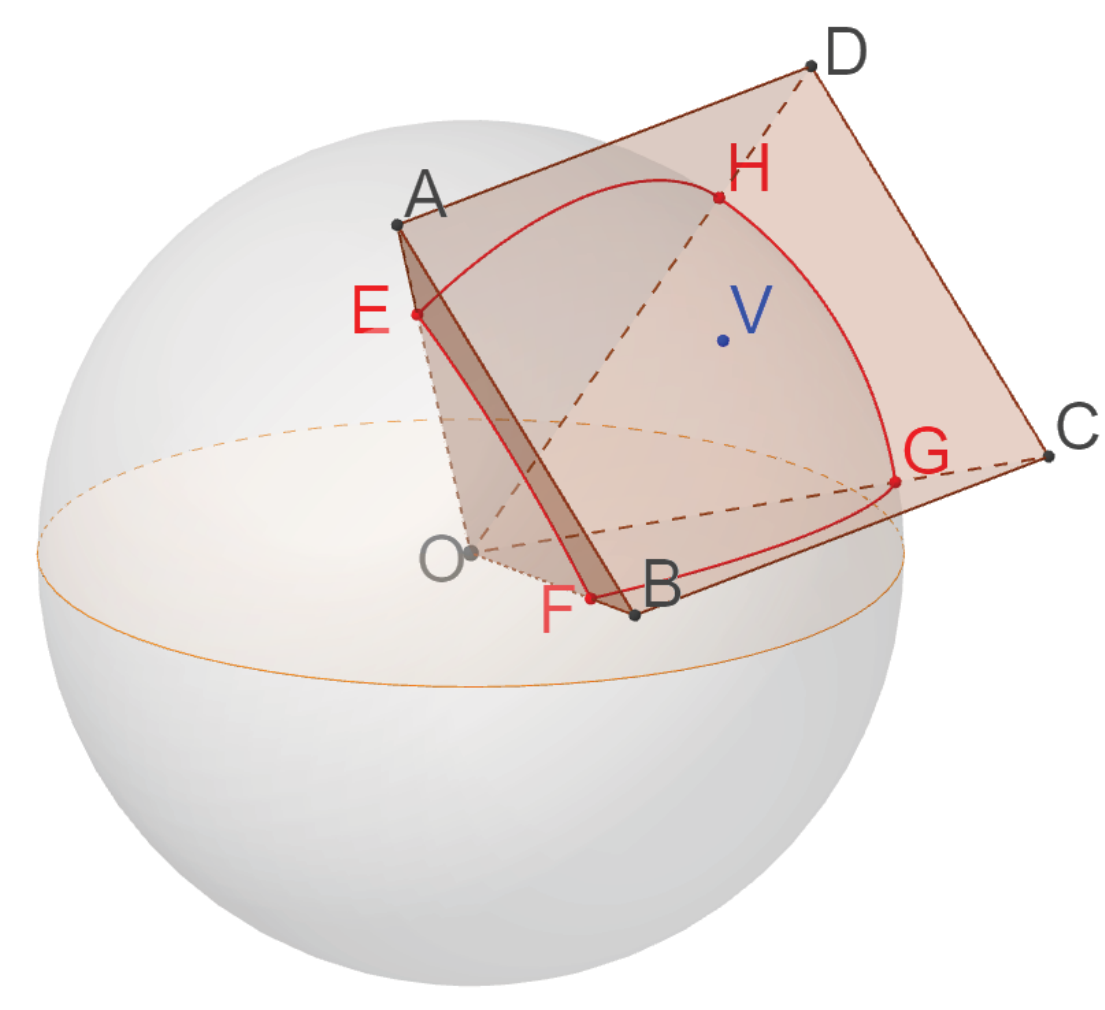}
  \end{minipage}
}
  \caption{(a) The FoVs of the HMD are $\alpha$ horizontally and $\beta$ vertically; (b) mapping the viewport plane $ABCD$ onto the surface $EFGH$ on the sphere.}
\label{fig:surface}\vspace*{-0.4cm}
\end{figure}

Actually, the image plane of the HMD can be considered as a 2-D plane with the region of the plane constrained by the FoV of the HMD. As an example, Fig. \ref{fig:surface}(a) shows the image plane $ABCD$ of an HMD, whose horizontal and vertical FoVs are $\alpha$ and $\beta$ respectively. Note that the plane $ABCD$ is tangent to the sphere at the user's viewpoint $\mathcal{V}=(\theta,\varphi)$. In Fig. \ref{fig:surface}(b), the surface $EFGH$ represents the result of mapping plane $ABCD$ onto the sphere, i.e., the user's viewport. Given $\alpha$, $\beta$ and the user's viewpoint $\mathcal{V}$, the plane $ABCD$ and the surface $EFGH$ are uniquely determined. Therefore, our problem becomes to find the viewport tile region in the 2-D projection plane based on the given plane $ABCD$ and surface $EFGH$.

\begin{figure}[!t]
 \centering
  \includegraphics[width=0.45\linewidth]{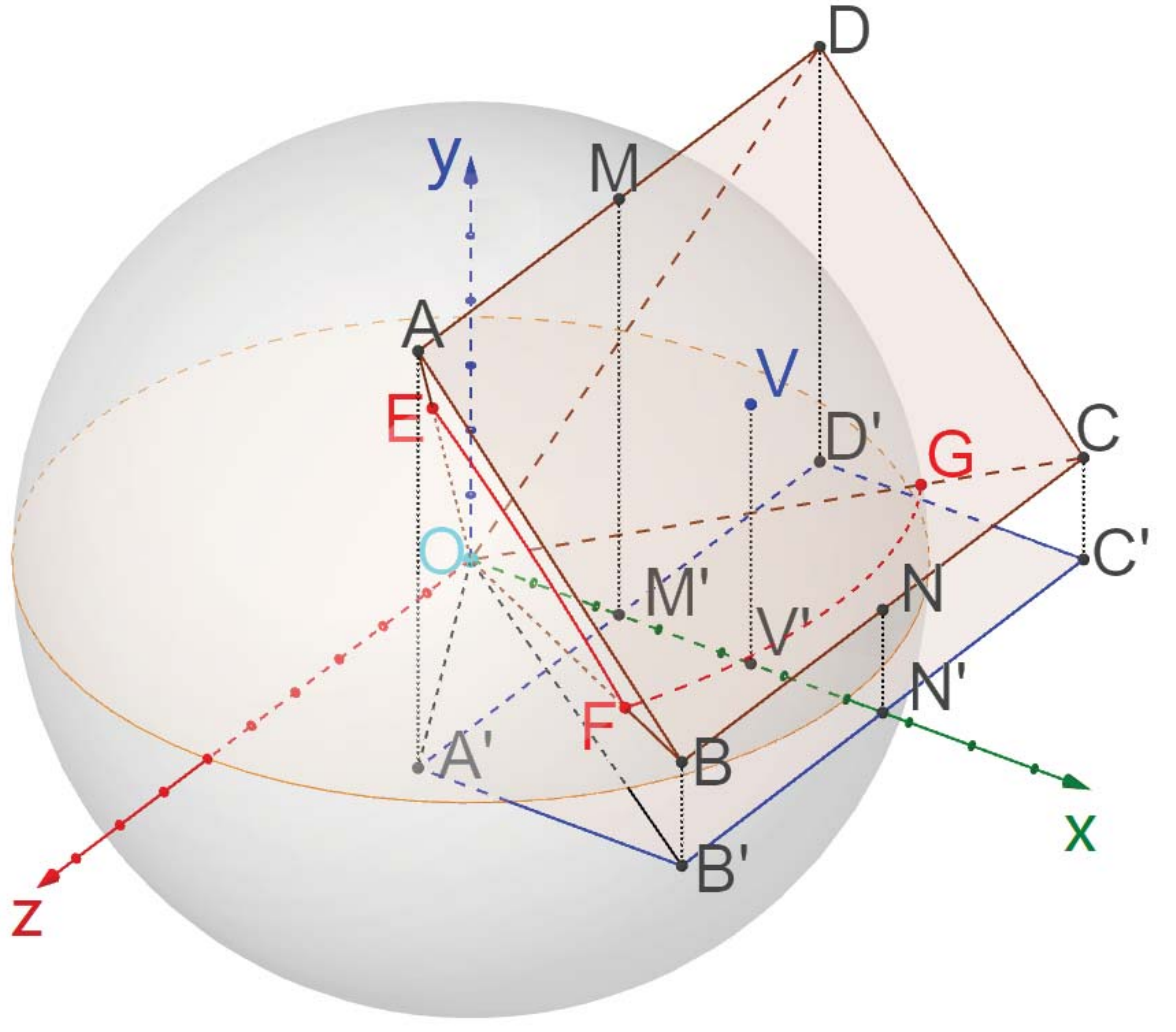}
  \caption{Calculation of the boundary of the user's viewport.}
  \label{fig:angle}\vspace*{-0.2cm}
\end{figure}

 \begin{figure}[!t]
    \centering
    \subfigure[]{ \label{fig_a}
    \includegraphics[width=1.75in]{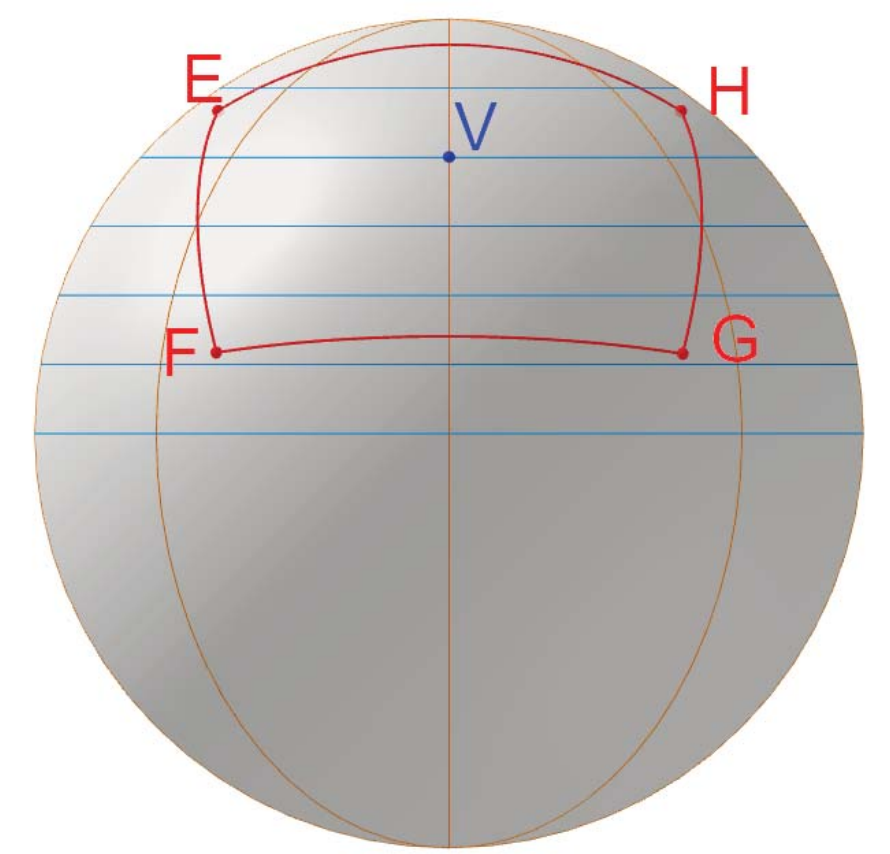} }
    \subfigure[]{ \label{fig_b}
    \includegraphics[width=1.7in]{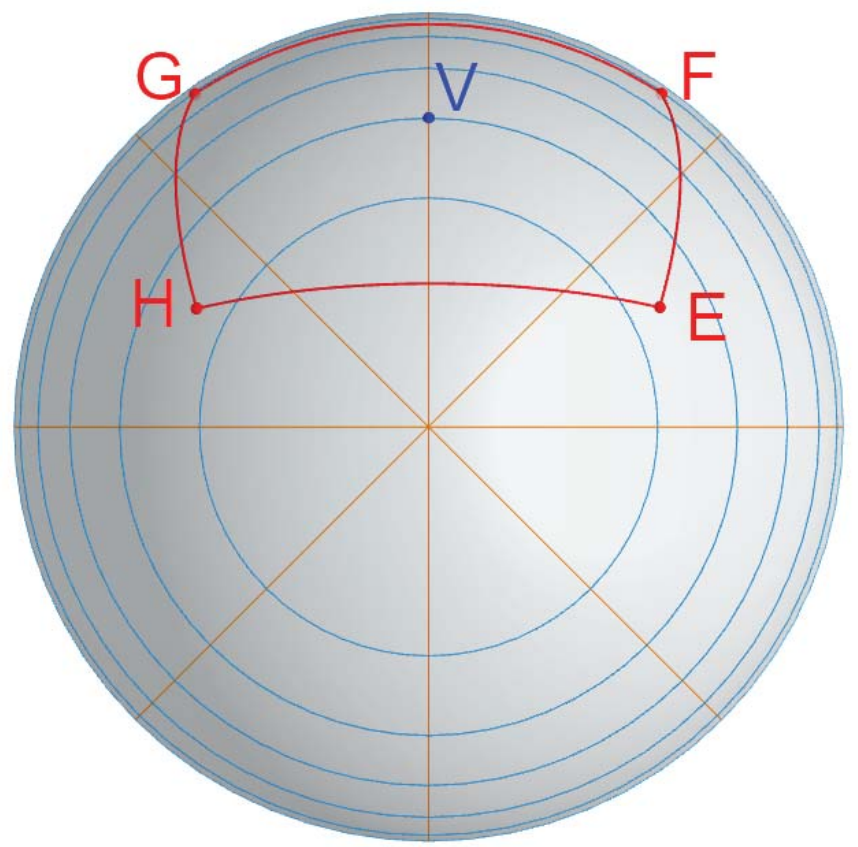} }
    \subfigure[]{ \label{fig_c}
    \includegraphics[width=1.7in]{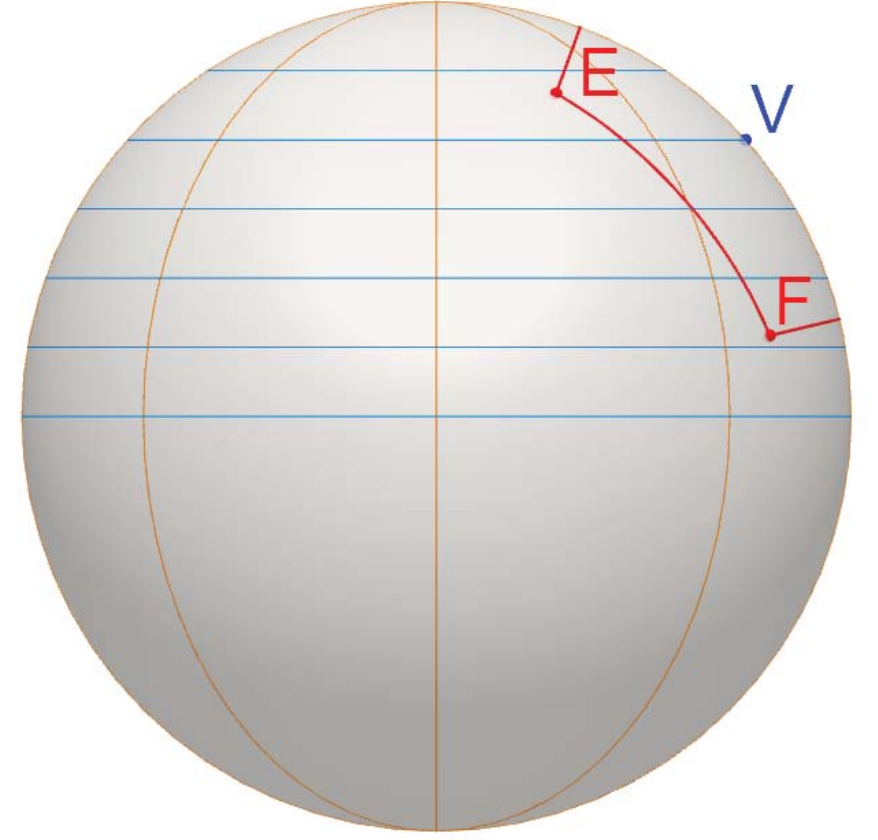} }
    \caption{ Three-view drawing of the viewport, where red lines represent the viewport boundary, yellow lines represent the longitude lines and blue lines are the latitude lines. (a) Frontal view; (b) vertical view; (c) lateral view.}\vspace*{-0.4cm}
    \label{fig:three-view}
    \end{figure}

To find the viewport tile region, intuitively, we only need to know the bounding box rather than the exact shape of the viewport. As seen in Fig. \ref{fig:viewport}, we can straightforwardly figure out the yellow color region as long as we know the bounding box of the viewport indicated by the blue rectangle. Considering that the bounding box of the viewport is uniquely determined by four extreme points, i.e., the westernmost, easternmost, northernmost and southernmost points on the viewport, we propose the following propositions to find these four points.

\begin{proposition}
\label{eq:lem}
If the user's viewpoint $\mathcal{V}=(\theta,\varphi)$ is located in the northern hemisphere, then point $E$ is the westernmost point of  the viewport and point $H$ is the easternmost point of the viewport. Otherwise, point $F$ and $G$ are respectively the westernmost and easternmost point of the viewport.
\end{proposition}

\begin{IEEEproof}
Fig. \ref{fig:angle} illustrates the geometric position of the viewport in the spherical space when the viewpoint is located in the northern hemisphere. Plane $A'B'C'D'$ is generated by projecting the plane $ABCD$ onto the plane $xOz$ vertically, arc $EF$ is in the plane $OAB$ and $EF$ must be projected into the triangle $OA'B'$ vertically. Since the longitude at a point would be equal to the angle between a vertical north-south plane through that point and the plane of the longitude $0^\circ$, denoted as $yOz$ , the westernmost point in the triangle $OA'B'$ is located on line segment $OA'$, which means the point $E$ is the westernmost point of the viewport. Known by symmetry, the point $H$ is the easternmost point of the viewport. By symmetry, if the viewpoint is at the southern hemisphere, the points $F$ and $G$ are the westernmost and easternmost points, respectively.
\end{IEEEproof}

The boundary of the viewport is also displayed in Fig. \ref{fig:three-view}, in which the boundary crosses different latitudes and longitudes. Similar to Proposition 1, we can derive Proposition 2:

\begin{proposition}
There are four cases for the location of the boundary points:
    1) if the latitude $\theta$ of the user's viewpoint $\mathcal{V}=(\theta,\varphi)$ satisfies $-90^\circ \le \theta \le (\frac{\beta}{2}-90^\circ)$, then the latitude of the southernmost point is $-90^{\circ}$ and the northernmost points are $E$ and $H$; 2) if the latitude $\theta$ satisfies $(90^\circ-\frac{\beta}{2}) \le \theta \le 90^\circ$, then the latitude of the northernmost point is $90^{\circ}$ and the southernmost points are $F$ and $G$; 3) if the latitude $\theta$ satisfies $-\frac{\beta}{2} < \theta < \frac{\beta}{2}$, then the southernmost point is located on arc $GF$ and between points $G$ and $F$, and the northernmost point is located on arc $EH$ and between point $E$ and $H$; 4) otherwise, if the viewpoint is in the northern hemisphere, then the northernmost point is on arc $EH$ and between points $E$ and $H$, and point $G$ and $F$ both are the southernmost points; if the viewpoint is in the southern hemisphere, the southernmost point is on arc $GF$ and between points $G$ and $F$, and points $E$ and $H$ both are the northernmost points. In addition, when the northernmost or southernmost point is between any two points, its latitude value is $\theta + \frac{\beta}{2}$ or $\theta - \frac{\beta}{2}$.
\end{proposition}

Once we obtain the boundary points in the spherical space, the next critical step is to calculate their longitude and latitude values so as to find the viewpoint tile region. For simplicity, we assume that the viewpoint $\mathcal{V}$ satisfies $\theta > \frac{\beta}{2}$ and $\varphi = 90^\circ$, since the longitude value $\varphi$ of the viewpoint has no effect on the longitude difference between the viewpoint and any point on the viewport. Hence, the longitudes of points $E$ and $H$ can be calculated when we achieve their longitude difference $\Delta $ from the latitude and longitude of viewpoint $\mathcal{V}$. The longitude difference $\Delta$ can be calculated as follows.

As shown in Fig. \ref{fig:angle}, we should get the angle $\delta$ between the $z$-axis and the line segment $OA'$ first for calculating the longitude of point $E$. The angel $\delta$ equals to the angle $\angle OA'D'$ and the distance from the viewpoint $\mathcal{V}$ to $O$ equals to the radius $R$ of the sphere. Then the distance from the viewpoint to line segment $AD$ and $AB$ is $R\textrm{tan}(\beta/2)$ and $R\textrm{tan}(\alpha/2)$, respectively. $M'$ is the point at which point $M$, midpoint of $AD$, is projected perpendicularly onto the plane $xOz$, so we have $VM = R\textrm{tan}(\beta/2), A'M'= R\textrm{tan}(\alpha/2)$, and the angle between the line segment $MV$ and the plane $xOz$ is $90^\circ -\theta$, with the angle $\angle VOV' = \theta$. Therefore, we have
\begin{equation}
V'M' = VM \textrm{cos}(90^\circ - \theta) = R\textrm{tan}(\beta/2) \textrm{cos}(90^\circ - \theta),
 \end{equation}
and the length of $OM'$ can be calculated by
    \begin{equation}
        OM' =OV' - V'M'= R\textrm{cos}\theta - R\textrm{tan}(\beta/2) \textrm{cos} (90^\circ -\theta).
    \end{equation}
    With the length of $A'M'$ and $OM'$, it is known that
    \begin{align}
        \delta & = \angle OA'D'  = \textrm{arctan}\bigg( \frac{OM'}{A'M'}\bigg) \nonumber \\
        & = \textrm{arctan}\bigg( \frac{R\textrm{cos}\theta - R\textrm{tan}(\beta/2) \textrm{cos} (90^\circ -\theta)}{R\textrm{tan}(\alpha/2)}\bigg),
    \end{align}
    and
    \begin{align}
        \Delta/2 & = 90^{\circ} - \delta \nonumber \\
        & =  90^{\circ} - \textrm{arctan}\bigg( \frac{R\textrm{cos}\theta - R\textrm{tan}(\beta/2) \textrm{cos} (90^\circ -\theta)}{R\textrm{tan}(\alpha/2)}\bigg),
    \end{align}
for the case where $R\textrm{cos}\theta - R\textrm{tan}(\beta/2) \textrm{cos} (90^\circ -\theta) \ge 0$; otherwise, the viewport covers the north pole and we thus define $\Delta/2=180^\circ$.

For any given viewpoint $\mathcal{V}=(\theta,\varphi)$ located in the northern hemisphere, the westernmost longitude $\varphi_{\textrm{wm}}$ and easternmost longitude $\varphi_{\textrm{em}}$ of the viewport are given by
    \begin{equation}
        \varphi_{\textrm{wm}} = [\varphi - \Delta/2]_{l_0},
    \end{equation}
    \begin{equation}
        \varphi_{\textrm{em}} = [\varphi + \Delta/2]_{l_0},
    \end{equation}
    where the function
    \begin{align}\label{eq:l_0}
        [x]_{l_0} = \left \{ \begin{array}{ll}
        x + 360^\circ  & ,\textrm{if $x < -180^\circ$}\\
        x - 360^\circ & ,\textrm{if $x > 180^\circ$} \end{array} \right.
    \end{align}
ensures that the term always takes value between $[-180^\circ,180^\circ]$.

Similarly, as seen in Fig. \ref{fig:angle}, point $N$ is the midpoint of $BC$, and point $N'$ is the point at which $N$ is vertically projected onto the plane $xOz$. Since $\theta > \frac{\beta}{2}$, the southernmost point of the viewport is point $F$ (or $G$) with its latitude equal to $\angle BOB'$ (or $\angle COC'$). $\overrightarrow{OV}$ is the normal vector of plane $ABCD$, $\angle N'OV = \theta$, so the angle between plane $xOz$ and plane $ABCD$ is $90^\circ - \theta$. And recall that the line segment $VM = R \textrm{tan}(\beta/2)$, so the length of $V'N'$ is $R \textrm{tan}(\beta/2) \textrm{cos}(90^\circ - \theta)$, hence we have
\begin{equation}
ON' = OV' + V'N'=R\textrm{cos}\theta + R \textrm{tan}(\beta/2) \textrm{cos}(90^\circ - \theta).
\end{equation}
With $BN = B'N' = R \textrm{tan}(\alpha/2)$, the length of $OB'$ is given by
    \begin{align}
            OB' &=\sqrt{(ON')^2 + (B'N')^2} \\
            &= \sqrt{[R\textrm{cos}\theta + R \textrm{tan}(\beta/2) \textrm{cos}(90^\circ - \theta)]^2 + [R \textrm{tan}(\alpha/2)]^2}. \nonumber
    \end{align}
And with the $VV' = R\textrm{sin}\theta$, the length of $BB'$ can be calculated as
    \begin{align}
         BB' & = VV' - VN\textrm{sin}(90^\circ - \theta) \\
        & = R\textrm{sin}\theta - R \textrm{tan}(\beta/2) \textrm{sin}(90^\circ - \theta). \nonumber
    \end{align}
Finally, we have
    \begin{align}
         \angle BOB' &= \textrm{arctan} \biggl (\frac{BB'}{OB'} \biggr) \\
        & = \textrm{arctan} \biggl(\frac{ R\textrm{sin}\theta - R \textrm{tan}(\beta/2) \textrm{sin}(90^\circ - \theta)}{\sqrt{[R\textrm{cos}\theta + R \textrm{tan}(\beta/2) \textrm{cos}(90^\circ - \theta)]^2 + [R \textrm{tan}(\alpha/2)]^2}} \biggr). \nonumber
    \end{align}

Therefore, the latitude $\theta_{\textrm{sm}}$ of the southernmost point of the viewport is
    \begin{align}
        \theta_{\textrm{sm}} = \left \{ \begin{array}{ll}
        -90^\circ & \textrm{if } -90^\circ \le \theta \le (\frac{\beta}{2}-90^\circ) \\
        \theta - \frac{\beta}{2} & \textrm{if } (\frac{\beta}{2}-90^\circ) < \theta \le \frac{\beta}{2} \\
        \theta' & \textrm{otherwise}, \end{array} \right.
    \end{align}
    where
    \begin{equation}
        \theta' = \textrm{arctan} \biggl (\frac{ R\textrm{sin}\theta - R \textrm{tan}(\beta/2) \textrm{sin}(90^\circ - \theta)}{\sqrt{[R\textrm{cos}\theta + R \textrm{tan}(\beta/2) \textrm{cos}(90^\circ - \theta)]^2 + [R \textrm{tan}(\alpha/2)]^2}} \biggr),
    \end{equation}
and similarly the latitude $\theta_{\textrm{nm}}$ of the northernmost point on the viewport can be derived as
    \begin{align}
        \theta_{\textrm{nm}} = \left \{ \begin{array}{ll}
        90^\circ & ,\textrm{if }(90^\circ-\frac{\beta}{2}) \le \theta \le 90^\circ \\
        \theta + \frac{\beta}{2} & ,\textrm{if } -\frac{\beta}{2} \le \theta < (90^\circ-\frac{\beta}{2}) \\
        \theta'' & ,\textrm{otherwise}, \end{array} \right.
    \end{align}
    where
    \begin{equation}
        \theta'' = -\textrm{arctan} \biggl (\frac{- R\textrm{sin}\theta - R \textrm{tan}(\beta/2) \textrm{sin}(90^\circ + \theta)}{\sqrt{[R\textrm{cos}\theta + R \textrm{tan}(\beta/2) \textrm{cos}(90^\circ + \theta)]^2 + [R \textrm{tan}(\alpha/2)]^2}} \biggr ).
    \end{equation}

When we obtain the westernmost longitude $\varphi_{\textrm{wm}}$, the easternmost longitude $\varphi_{\textrm{em}}$,  the northernmost latitude $\theta_{\textrm{nm}}$, and the southernmost latitude $\theta_{\textrm{sm}}$ of user's viewport, the set of tiles in the viewport tile region can be determined. Specifically, assume that the rectangular planar video frame is divided into $M \times N$ tiles, with each tile crossing $180^\circ/M$ vertically and $360^\circ/N$ horizontally. Let $T(m,n)$ denote the tile located in the $m$-th row and the $n$-th column, with $ m \in [1,2,...,M]$ and $ n \in [1,2,...,N]$, as shown in Fig. \ref{fig:viewport}. In this way, the viewport tile region vertically ranges from the $m_0$-th to the $m_1$-th row, and horizontally ranges from the $n_0$-th to the $n_1$-th column, with
\begin{equation}\nonumber
\begin{aligned}
m_0=\left\lceil M-\frac{M\times (\theta_{\textrm{nm}}+ 90^\circ)}{180^\circ} \right\rceil,
m_1=\left\lceil M-\frac{M\times (\theta_{\textrm{sm}}+ 90^\circ)}{180^\circ} \right\rceil,
\end{aligned}
\end{equation}
\begin{equation}
\ \ n_0=\left\lceil \frac{N\times (\varphi_{\textrm{wm}}+ 180^\circ)}{360^\circ}\right\rceil,
n_1=\left\lceil \frac{N\times (\varphi_{\textrm{em}}+ 180^\circ)}{360^\circ} \right\rceil,
\end{equation}
where $\left\lceil \cdot \right\rceil$ represents the ceiling operation that maps the value inside the operation to the least integer greater than or equal to that value.

\subsection{Tile Visibility Probability}
If the proposed viewport prediction is absolutely accurate, the tiles in the viewport tile region $\mathcal{B}_v$ should be visible to the user with a probability of 1, and all the other tiles are invisible to the user. Since prediction errors can hardly be avoided, some tiles located outside the viewport tile region may be visible to the user. To calculate the visibility probability of those tiles, we first analyze the viewport prediction error.

\begin{figure}[!t]
    \centering
    \subfigure[]{ \label{fig_a}
    \includegraphics[width=3.075in]{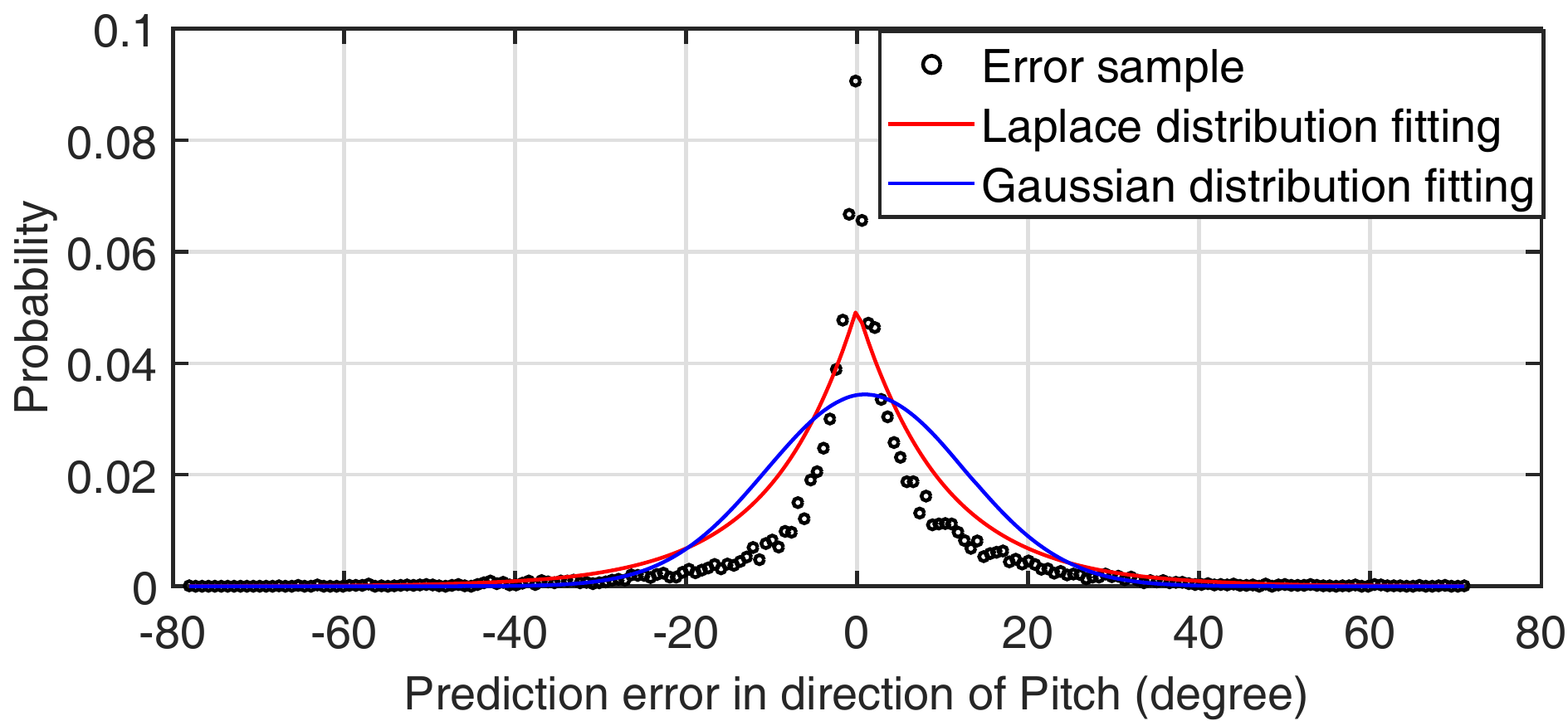} }
    \subfigure[]{ \label{fig_b}
    \includegraphics[width=3.075in]{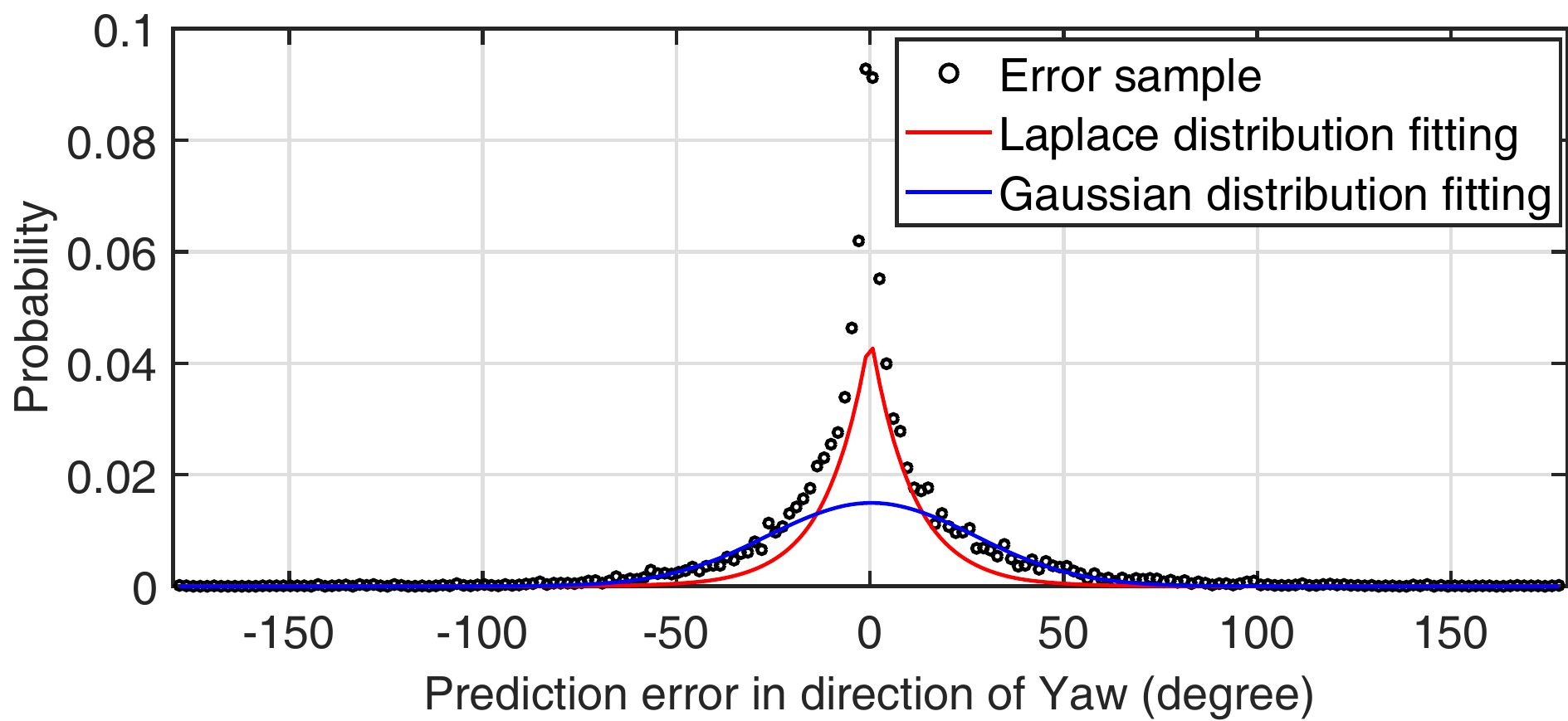} }
    \caption{Prediction error distribution of pitch and yaw angles.}
    \label{fig:error}\vspace*{-0.6cm}
 \end{figure}

The authors in the literature \cite{Steinbach2018, Xie2017} assumed that the prediction error of the head motion follows a Gaussian distribution. However, using the proposed viewport prediction model on the real head movement traces (to be detailed in Sec. \ref{sec:simulate}), we find that the Laplace distribution, rather than the Gaussian distribution, can approximate the distribution of the prediction error more accurately. This is illustrated in Fig. \ref{fig:error}, where we fit the probability density function of the prediction error with the Laplace and Gaussian distributions, respectively. Through running the Jarque-Bera test \cite{Jarque-Bera}, the hypothesis of Gaussian distribution for the prediction error data shown in Fig. \ref{fig:error} has been rejected at the $5\%$ significance level. In comparison, it can be seen that the Laplace distribution is more accurate. Therefore, in the rest of this paper, we assume the Laplace distribution for the prediction error of the pitch and yaw angles. Namely, we have
\begin{equation}
    {p_\theta  (\Delta \theta)} = \frac{1}{\lambda_{\theta}} e^{-\frac{|\Delta \theta|}{\lambda_{\theta}}},
    {p_\varphi (\Delta \varphi)}= \frac{1}{\lambda_{\varphi}} e^{-\frac{|\Delta  \varphi|}{\lambda_{\varphi}}},
\end{equation}
where $\Delta \theta$ and $\Delta \varphi$ denote the prediction errors for the pitch and yaw angles; the scale parameters $\lambda_{\theta}$ and $\lambda_{\varphi}$ can be learned from training data.


For any tile $T(m,n)$, we can calculate its upper, lower, left and right boundary, respectively, as follows.
\begin{equation}
\theta^{\text{upper}}_{m,n}=90^\circ - \frac{(m-1) \times 180^\circ}{M}, \theta^{\text{lower}}_{m,n}=90^\circ - \frac{m \times 180^\circ}{M}, \nonumber
\end{equation}
\begin{equation}
\varphi^{\text{left}}_{m,n}=-180^\circ + \frac{(n-1) \times 360^\circ}{N}, \varphi^{\text{right}}_{m,n}=-180^\circ + \frac{n \times 360^\circ}{N}.
\end{equation}
This tile is visible to the user if the following conditions on the prediction error are satisfied:
\begin{equation}\label{eq:errorBoundTheta}
\theta_{\textrm{nm}} + \Delta \theta \ge \theta^{\text{lower}}_{m,n},\ \theta_{\textrm{sm}} + \Delta \theta \le \theta^{\text{upper}}_{m,n},
\end{equation}
\begin{equation}\label{eq:errorBoundVarphi}
\varphi_{\textrm{em}} + \Delta \varphi \ge \varphi^{\text{left}}_{m,n},\ \varphi_{\textrm{wm}} + \Delta \varphi \le \varphi^{\text{right}}_{m,n},
\end{equation}
Eqs. (\ref{eq:errorBoundTheta}) and (\ref{eq:errorBoundVarphi}) together specify that given the predication errors $\Delta \theta$ and $\Delta \varphi$, the tile $T(m,n)$ is within the user's viewport if its northernmost latitude plus $\Delta \theta$ is still higher than the lower boundary of tile $T(m,n)$, its southernmost latitude plus $\Delta \theta$ is still lower than the upper boundary, its easternmost longitude plus $\Delta \varphi$ is still right to the left boundary, and its westernmost longitude plus $\Delta \varphi$ is still left to the right boundary. Assuming that the prediction errors of the pitch and yaw angles are independent, then the visibility probability of any tile $T(m,n)$ can be derived as:
\begin{align}\label{eq:Pvisible}
P_{m,n} \triangleq P_{m,n}^{\theta} \cdot P_{m,n}^{\varphi} =\int_{\max\{\theta^{\text{lower}}_{m,n}-\theta_{\textrm{nm}},-90^\circ\}}^{\min\{\theta^{\text{upper}}_{m,n}-\theta_{\textrm{sm}},90^\circ\}} p_\theta (\Delta \theta) d \Delta \theta \cdot \int_{[\varphi^{\text{left}}_{m,n}-\varphi_{\textrm{em}}]_{l_0}}^{[\varphi^{\text{right}}_{m,n}-\varphi_{\textrm{wm}}]_{l_0}} p_\varphi (\Delta \varphi) d \Delta \varphi,
\end{align}
where $P_{m,n}^\theta$ and $P_{m,n}^\varphi$ are defined for notational simplicity; the functions $\max\{\cdot, -90^\circ\}$ and $\min\{\cdot, 90^\circ\}$ in the lower and upper limits of the first integral map the value outside the range of $[-90^\circ,90^\circ]$ to $-90^\circ$ and $90^\circ$, respectively; and the function $[\cdot]_{l_0}$ follows the definition in Eq. (\ref{eq:l_0}) to ensure that the value is within the range of $[-180^\circ,180^\circ]$.

It is noted that $P_{m,n}, \forall T(m,n) \notin \mathcal{B}_v$ decreases and approaches zero when the tile $T(m,n)$ gradually deviates from $\mathcal{B}_v$. In other words, the visibility probability of any tile outside the viewport region depends on its distance to the viewport region, and becomes approximately zero if it is far away from the viewport region. For the sake of analysis, we introduce a probability threshold $\alpha$, according to which the tiles $T(m,n) \notin \mathcal{B}_v$ are further divided into \textit{marginal tiles} (if $P_{m,n} \ge \alpha$) and \textit{invisible tiles} (if $P_{m,n} < \alpha$). In this way, the tiles in a planar frame can be classified into three categories: \textit{viewport tiles}, \textit{marginal tiles} and \textit{invisible tiles}. The basic criterion to allocate rate between different types of tiles is as follows. Viewport tiles are allocated with an encoding rate higher than or equal to that of marginal tiles, since they will be displayed on the user's HMD with a much higher probability than marginal tiles. In addition, invisible tiles are allocated with the smallest encoding rate to promise the transmission of basic video quality. For a predicted user viewpoint $\mathcal{V}=(\theta,\varphi)$, viewport tiles are fixed, referring to the tiles within the corresponding viewport region $\mathcal{B}_v$, while the probability threshold $\alpha$ adjusts the number allocation between marginal and invisible tiles. As $\alpha$ decreases towards zero, the number of marginal tiles will be increased to be more tolerant to prediction error of the user's viewpoint, at the cost of increased transmission rate. In the next section, we will discuss the server-side optimal rate allocation strategy for these tiles among multiple users.


\section{Tile Rate Adaptation}
\label{sec:rate}
In this section, we first formulate the server-side rate adaptation problem, and then develop a steepest descent solution.

\subsection{Problem Formulation}
\begin{figure}[!t]
  \centering
    \includegraphics[width=0.6\linewidth]{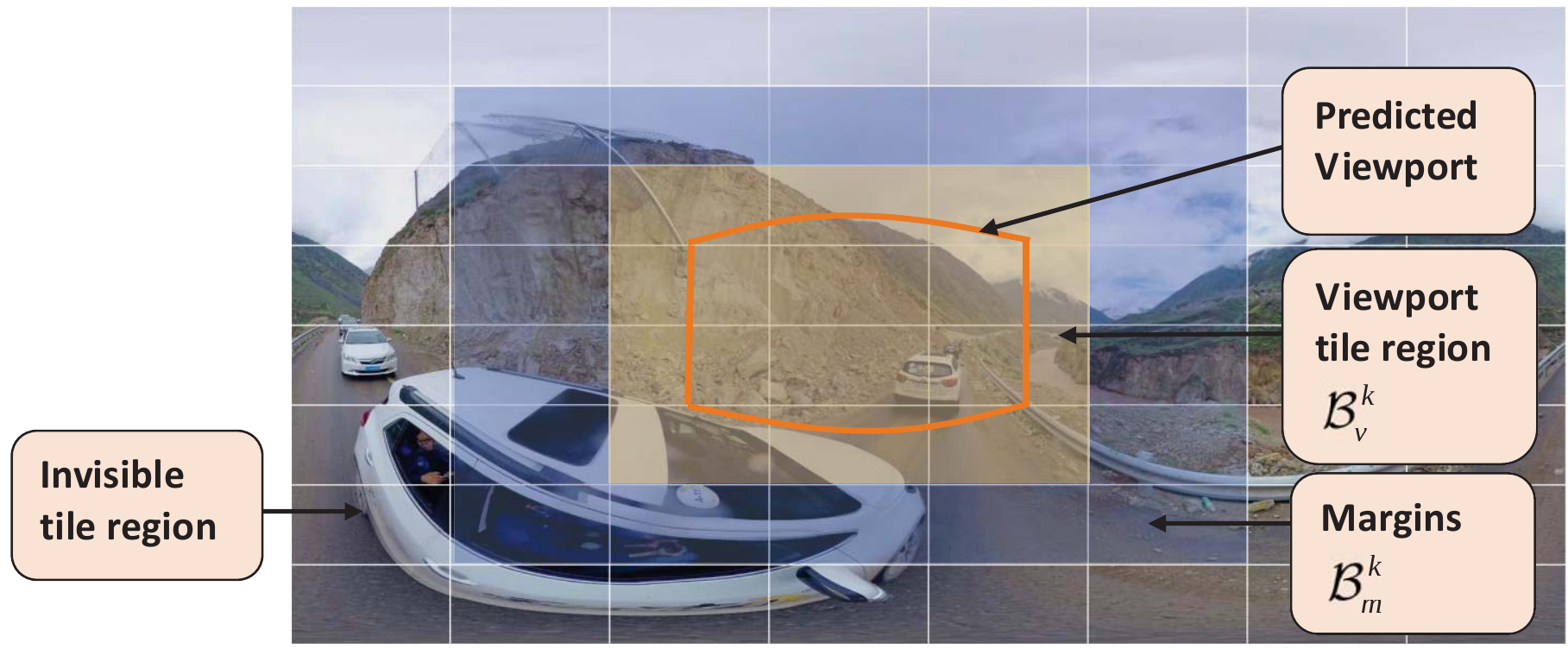}
  \caption{ Tile classification of a 2-D planar frame }
  \label{fig:FoV}\vspace*{-0.4cm}
\end{figure}

Assume that the 2-D planar video sequence is temporally divided into a set of segments of the same duration. Each frame within a segment is spatially divided into $M \times N$ tiles. Each tile is encoded into $L$ representations, with $\mathcal{R} = \{{R_1},{R_2},...,{R_L}\} $ denoting the set of encoding rates in an increasing order, i.e., $R_1<R_2<...<R_L$. Assume that the server sends the video simultaneously to a set $\mathcal{K}$ of users. The transmission capacity of the server and each user $k \in \mathcal{K}$ is denoted as $C^s$ and $C^k$, respectively. For each segment, let $\mathcal{B}_v^{k}$ denote the set of viewport tiles of user $k$, and $\mathcal{B}_m^{k}$ denote the corresponding set of marginal tiles. Further, let the combination $\mathcal{B}^{k}=\mathcal{B}_v^{k} \bigcup \mathcal{B}_m^{k}$ represent the set of \textit{visible tiles} that are visible to user $k$, and $\overline{\mathcal{B}}^{k}$, the supplementary set of $\mathcal{B}^{k}$, represent the set of \textit{invisible tiles}, as shown in Fig. \ref{fig:FoV}.

To evaluate the quality of the 360-degree video, the weighted-to-spherically-uniform PSNR (WS-PSNR) \cite{ISO2016} is widely used. It defines the weighted average distortion of the points in the plane as the 360-degree video distortion. The value of the weight (importance) depends on the position of the point. For instance, the weight equals one around the equator and decreases towards the poles, and finally approximates to zero at the poles.

To measure the video quality of each tile, we extend the pixel-based WS-PSNR definition to the tile-granularity and define the weighted distortion of the tile in the plane as its spherical distortion. The weight of the tile in the plane is proportional to its area of projection region in the spherical space. Namely, we have
\begin{equation}
\mathcal{D}_{m,n}^S=S_{m,n} \cdot D_{m,n},
\end{equation}
where $\mathcal{D}_{m,n}^S$ and $D_{m,n}$ are the distortion of the tile on the sphere and in the plane, respectively, and $S_{m,n}$ is the tile's area on the spherical surface that is given by
\begin{equation}
S_{m,n}=\int_{-\pi+ 2 \frac{n-1}{N} \pi}^{-\pi+ 2 \frac{n}{N} \pi} \int_{\frac{\pi}{2}- \frac{m}{M}\pi}^{\frac{\pi}{2}- \frac{m-1}{M}\pi} R^2 \cos \varphi d\varphi d\theta.
\end{equation}
Here, the distortion $D_{m,n}$ in the plane can be measured by the traditional rate-distortion model \cite{Stuhlmuller2000}:
\begin{equation}
\label{eq:distortion}
  D_{m,n}(R_{m,n})=\cfrac {\sigma} {R_{m,n}-R_0}+D_0,
\end{equation}
where $R_{m,n} $ is the encoding rate of the tile. The variables $\sigma $, $R_0 $ and $D_0 $ are the parameters of the R-D model, which can be fitted by the empirical data from trial encodings using nonlinear regression techniques.

As previously mentioned, the tiles in a planar frame are classified into three types: viewport, marginal and invisible tiles. The viewport tiles are streamed with high quality, while the marginal tiles are transmitted with moderate video quality. As for the invisible tiles, they are transmitted at the lowest rate $R_1$ to save transmission capacity.

For each video segment, the server seeks optimal transmission rates for the viewport and marginal tiles, aiming at minimizing the overall video distortion perceived by the users, subject to both the server and user rate constraints. Mathematically, the server-side rate adaptation problem can be formulated as the following problem \textbf{P1}:
\begin{subequations}\label{eq:P1}
\begin{align}
\textbf{P1:} & \ \mathop {\min }\limits_{\bf{R}} \;   \frac{1}{|\mathcal{K}|}\sum\limits_{k \in \mathcal{K}} \frac{1}{\sum_{T(m,n) \in \mathcal{B}^{k}} S_{m,n}} \cdot  \biggl [  \sum\limits_{T(m,n) \in \mathcal{B}^{k}} S_{m,n} \cdot D^k_{m,n}({R^k_{m,n}}) P^k_{m,n} \\
&  \quad \quad \quad  \quad \quad \quad  \quad \quad \quad  \quad \quad \quad  \quad \quad \quad  + \omega S_{\hat m,\hat n} {D^k_{\hat m,\hat n}\biggl (\mathop {\min }\limits_{_{T(m,n) \in {\cal B}^{k}_m}}{R^k_{m,n}}\biggr )} {P^k_{\hat m, \hat n}} \biggr ] \nonumber\\
\text{s.t.}\  & \quad \sum\limits_{k \in \mathcal{K}}{\sum\limits_{T(m,n) \in {\mathcal{B}^{k}}}} R^k_{m,n} \; \le C^s, \\
& \  {\sum\limits_{T(m,n) \in {\mathcal{B}^{k}}}} R^k_{m,n} \le C^k, \; \forall k \in \mathcal{K}, \\
& \  R^k_{m,n} \in \left\{ {{R_1},{R_2},...,{R_L}} \right\}, \; \forall T(m,n) \in \mathcal{B}^{k},\forall k \in \mathcal{K}, \\
& \  {R^k_{m,n} } = R_1, \; \forall T(m,n) \in \overline{\mathcal{B}}^{k},\forall k \in \mathcal{K}, \\
& \  {R^k_{m,n} } = {R^k_{m',n'} }, \; \forall T(m,n), T(m',n') \in {\cal B}^{k}_v, \forall k \in \mathcal{K}, \\
& \  {R^k_{m,n} } \le {R^k_{m',n'} }, \; \forall T(m,n) \in  {\cal B}^{k}_m, \forall T(m',n') \in  {\cal B}^{k}_v, \forall k \in \mathcal{K}.
\end{align}
\end{subequations}
In the objective function in Eq. (\ref{eq:P1}a), the first term defines the expected video quality in terms of the weighted-to-spherically-uniform MSE perceived by all the users, which is decreasing and strictly convex with respect to the video rate. The second term in the objective function is introduced to avoid uncomfortable degradation in QoE when the user switches the viewing direction from the predicted viewport to the margin area, where $\omega$ is a weight and $(\hat m, \hat n) =\arg \min_{T(m,n) \in \mathcal{B}^k_m} R^k_{m,n}$. Constraints in Eq. (\ref{eq:P1}d) define the optional quality levels for each visible tile. Constraints in Eq. (\ref{eq:P1}e) set the quality level of all the invisible tiles to $R_1$. Constraints in Eq. (\ref{eq:P1}f) specify that all the tiles in the viewport tile region should have the same quality level. Constraints in Eq. (\ref{eq:P1}g) ensure that the quality level in the margin area would not be higher than that in the viewport region. 

From Eq. (\ref{eq:distortion}), we know that $D_{m,n}(R_{m,n})$ is convex since $D_{m,n}''(R_{m,n})>0$ when $R_{m,n}>R_0$. Given that pointwise maximum preserves convexity \cite{Boyd}, we have that $\underset{T(m,n)\in \mathcal{B}_m^k}{\min} R_{m,n}^k=-\underset{T(m,n)\in \mathcal{B}_m^k}{\max} \left(-R_{m,n}^k\right)$ is concave. Then, based on the property of scalar composition that preserves convexity of the outer function if the inner function is concave \cite{Boyd}, $D_{\hat m, \hat n}^k\left(\underset{T(m,n)\in \mathcal{B}_m^k}{\min} R_{m,n}^k\right)$ is convex. Hence, the objective function of problem \textbf{P1} in Eq. (\ref{eq:P1}a) is a nonnegative weighted sum of convex functions, which is still convex.

\subsection{Steepest Descent Solution}
The problem \textbf{P1} is a nonlinear discrete optimization problem which is in general NP-hard. To solve this problem, an intuitive way is to search all possible choices for the variables. This may causes the computational complexity to increase exponentially. In contrast, a steepest descent algorithm can efficiently reduce the computational complexity to achieve an optimal or near-optimal solution by iteratively searching the steepest descent direction from which the variables are updated under all the constraints. In the following, based on the steepest descent approach, we propose an algorithm, of which the feasible starting point is achieved from the relaxed convex optimization problem, to solve the problem \textbf{P1} in polynomial time complexity.

Let $Q(\mathbf{R})$ denote the objective function of problem \textbf{P1}, where the vector $\mathbf{R} = \{R^k_{m,n}|\forall k \in \mathcal{K}, \forall T(m,n) \in \mathcal{B}^{k}\cup \overline{\mathcal{B}}^{k} \}$ \normalsize represents the bitrates of all tiles that are allocated for each user, and $B(\mathbf{R})$ denotes the consumed transmission rate of all users in $\mathcal{K}$, i.e., $B(\mathbf{R}) = \sum_{k \in \mathcal{K}}{\sum_{T(m,n) \in {\mathcal{B}^{k}\cup \overline{\mathcal{B}}^{k}}} } R^k_{m,n}$. Also, denote $\mathbf{R}^+_j$ as an operation that increases the rate value $R_u \in \mathcal{R}$ to the nearest higher rate value $R_{u+1} \in \mathcal{R}$ for the $j$-th element of the vector $\mathbf{R}$ while the other elements remain unchanged. Therefore, the slope in the $j$-th direction at $\mathbf{R}$ can be defined as
\begin{equation}\label{eq:slop}
    s_j(\mathbf{R}) = - \frac{Q(\mathbf{R}^+_j)-Q(\mathbf{R})}{B(\mathbf{R}^+_j)-B(\mathbf{R})}.
\end{equation}
The coordinate-wise steepest descent algorithm iteratively searches for the index $j^*$ with the steepest descent direction, i.e., the index $j^*$ that achieves the largest ratio of the distortion reduction to the rate increment for an increment in the bitrate by a single step within the available encoding rate set $\mathcal{R}$. If the constraints in Eqs. (\ref{eq:P1}b)-(\ref{eq:P1}g) of \textbf{P1} are still met after executing the operation $\mathbf{R}^+_{j^*}$, $\mathbf{R}$ will be updated in accordance with $\mathbf{R}^+_{j^*}$. To ensure the optimality of the solution, three optimality conditions for the $(Q(\mathbf{R}), B(\mathbf{R}))$ pair should be satisfied \cite{Sermadevi2004}:

1) \emph{Cross-over condition:} We define $Q^{R_u}_j(B)$ as the lower convex hull of achievable solutions with the $j$-th element fixed to $R_u$. If for each $j$ and $R_u<R_v,\ \forall R_u,R_v \in \mathcal{R}$, we have $Q^{R_v}_j(B)<Q^{R_u}_j(B)$ for a sufficiently large $B(\mathbf{R})$. Furthermore, if there exists some $B_0$ such that $Q^{R_v}_j(B) \le Q^{R_u}_j(B)$, we have $Q^{R_v}_j(B) < Q^{R_u}_j(B), \; \forall B(\mathbf{R}) > B_0$.

2) \emph{Cross-over ordering condition:} We define the cross-over bandwidth $B^c(R_u,R_v,j)$ between two curves $Q^{R_u}_j(B)$ and $Q^{R_v}_j(B)$ as the smallest consumed transmission rate such that $Q^{R_v}_j(B) \le Q^{R_u}_j(B)$ if $R_u<R_v$. Assume that $B^c(R_u,R_v,j)$ is not achievable, i.e., $B^c(R_u,R_v,j) \ne B(\mathbf{R}), \forall R^k_{m,n} \in \mathcal{R}$, where $R^k_{m,n}$ is the element of vector $\bf{R}$. If for each $j$ and $R_u<R_v<R_w$, $B^c(R_v,R_w,j)$ is greater than the consumed transmission rate of the first achievable solution on $Q^{R_v}_j(B)$ with the consumed transmission rate greater than or equal to $B^c(R_u,R_v,j)$.

3) \emph{Reachability condition:} If for any $j$ and $R_u$, the difference between the two vectors corresponding to two adjacent achievable solutions on $Q^{R_u}_j(B)$ is exactly one index. In addition, given a vector $\mathbf{R}$ corresponding to the highest consumed transmission rate achievable solution smaller than or equal to $B^c(R_u,R_{u+1},j)$ on $Q^{R_u}_j(\mathbf{R})$, the vector $\mathbf{R}^+_j$ gives the lowest achievable consumed transmission rate solution on $Q^{R_{u+1}}_j(B)$ with the consumed transmission rate greater than or equal to $B^c(R_u,R_{u+1},j)$.

\begin{figure}[!t]
  \centering
    \includegraphics[width=0.55\linewidth]{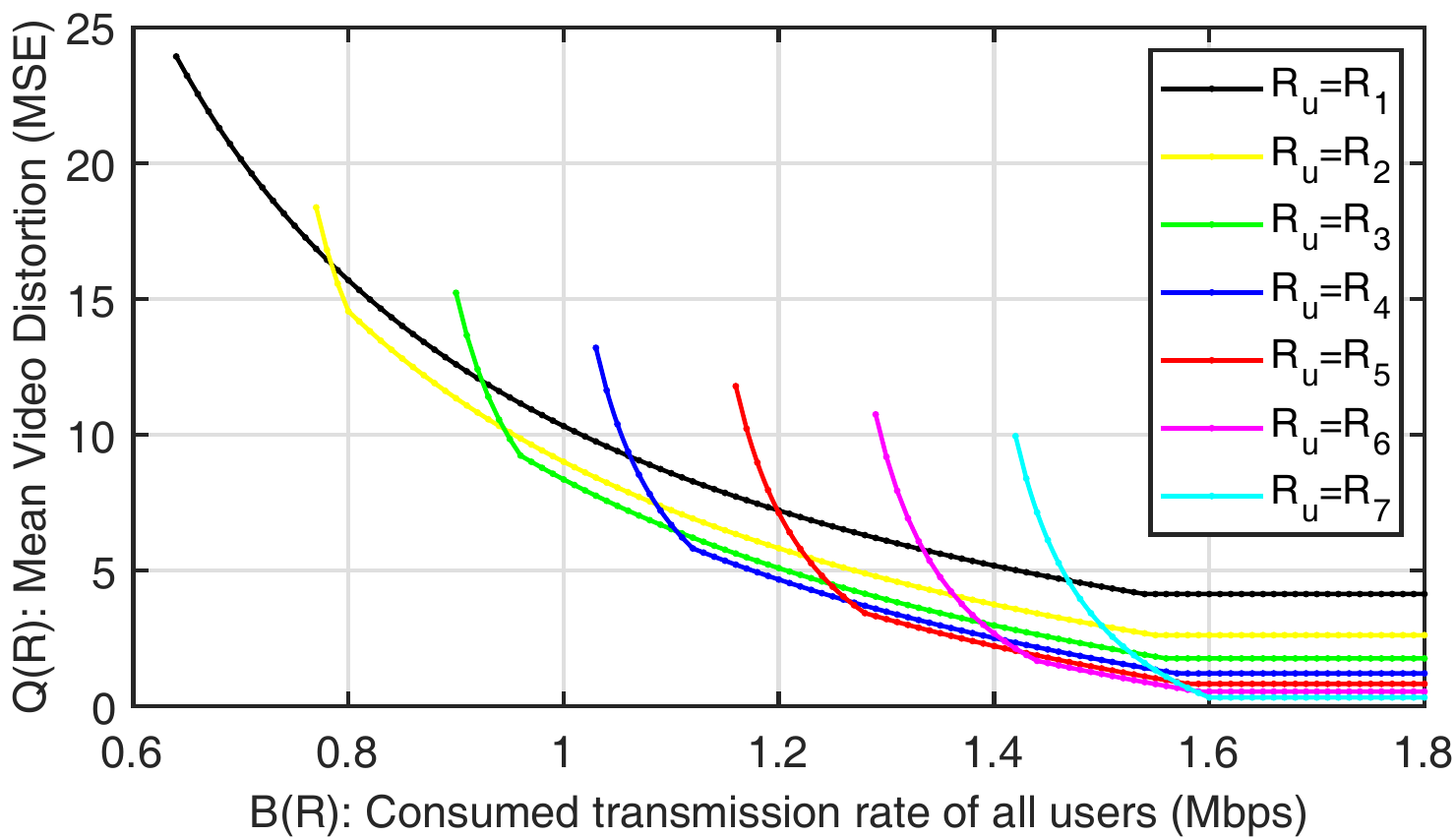}
  \caption{An example of the lower convex hull of $Q^{R_u}_j(B)$ with the $j$-th element fixed to different values of $R_u$.}
  \label{fig:Q-B}\vspace*{-0.6cm}
\end{figure}

It is proved in \cite{Sermadevi2004} that the steepest descent algorithm is able to find all achievable $(Q(\mathbf{R}),$$ B(\mathbf{R}))$ values on the lower convex hull with the consumed transmission rate $B(\mathbf{R})$ satisfying the constraints of problem \textbf{P1} if the cross-over, cross-over ordering and reachability conditions are satisfied. As an illustrative example, Fig. \ref{fig:Q-B} shows that the $(Q(\mathbf{R}), B(\mathbf{R}))$ pairs in the proposed optimization problem \textbf{P1} satisfy the \textit{cross-over} and \textit{cross-over ordering} conditions. Specifically, for the increasing encoding bitrates $R_1 < R_2 < \cdots< R_7$, we have $Q^{R_7}_j(B) < Q^{R_6}_j(B) < \cdots < Q^{R_1}_j(B)$ for $B(\mathbf{R})>1.6$ Mbps, which indicates that the \emph{cross-over} condition holds. Also, the \emph{cross-over ordering} condition is verified since $B^c(R_1,R_2,j) < B^c(R_2,R_3,j) < \cdots < B^c(R_6, R_7, j) $. The \emph{reachability condition} may not be guaranteed as there are too many directions in which $\mathbf{R}^+_j$ gives the lowest achievable solution. However, experimental results in Fig. \ref{fig:global} (please refer to Section \ref{sec:simulate}) still suggest that the solution achieved by the steepest descent algorithm in Algorithm 1 is close to the optimal solution obtained by the global search approach.

\begin{algorithm}[!t]
\caption{Steepest descent algorithm with initialization based on the continuous relaxation of problem \textbf{P1}}
\begin{algorithmic}[1]
\STATE \textbf{Initialization:}
\STATE Convert the original problem \textbf{P1} into its continuous relaxation problem \textbf{P2}, by relaxing the discrete rate constraint in Eq. (\ref{eq:P1}d) as $R^k_{m,n} \in [R_1,R_L]$.
\STATE Obtain the optimal solution $\widetilde{\bf{R}} := \{\widetilde{R}^k_{m,n}| \forall k,m,n\}$ to the continuous relaxation problem \textbf{P2} by the standard convex optimization technique.
\STATE Initialize the feasible starting point as $\mathbf{R} :=\{\lfloor \widetilde{R}^k_{m,n}\rfloor_{\mathcal{R}} | \forall k,m,n\}$.
\STATE Let the \emph{active searching set} $\mathcal{A}$ include the indices $\{j\}$ to which the corresponding elements in the rate vector $\mathbf{R}$ can be executed by the operation $\mathbf{R}^+_{j}$ while the constraints in Eqs. (\ref{eq:P1}b)-(\ref{eq:P1}g) are still satisfied.
\STATE
\STATE \textbf{Steepest descent search:}
    \WHILE {$\mathcal{A}$ is nonempty}
    \STATE Compute $s_j(\mathbf{R})$ for all $j \in \mathcal{A}$ according to Eq. (\ref{eq:slop}).
    \STATE Obtain the index $j^* :=\arg \max_{j \in \mathcal{A}} s_j(\mathbf{R})$ with the corresponding $T^*(m,n)$ and $k^*$.
    \IF{$T^*(m,n) \in \mathcal{B}^{k^*}_{\mathcal{V}}$}
        \FOR {each index $i$ corresponding to $ k = k^*, T(m,n) \in \mathcal{B}^{k^*}_{\mathcal{V}}$}
            \STATE $\mathbf{R}' := \mathbf{R}^+_i$
        \ENDFOR
    \ELSIF{$T^*(m,n) \in \mathcal{B}^{k^*}_{m}$}
        \STATE $\mathbf{R}' := \mathbf{R}^+_{j^*}$
    \ENDIF
    \IF{$\mathbf{R}' $ satisfy the constrains in Eqs. (\ref{eq:P1}b)-(\ref{eq:P1}g)}
        \STATE $\mathbf{R}^* := \mathbf{R}'$
    \ELSE
        \STATE Remove $j^*$ from $\mathcal{A}$.
    \ENDIF
    \ENDWHILE
\RETURN $\mathbf{R}^* $
\end{algorithmic}
\end{algorithm}

The detailed implementation process of the proposed steepest descent algorithm is shown in Algorithm 1. To avoid the steepest descent search dropping into a local optimum, it is important to select a feasible starting point in proximity to the global optimal solution. To this end, we first convert the original problem \textbf{P1} into its continuous relaxation problem \textbf{P2}, by relaxing the discrete rate constraint in Eq. (\ref{eq:P1}d) as $R^k_{m,n} \in [R_1,R_L]$. We then solve the continuous relaxation problem \textbf{P2} by the standard convex optimization technique to obtain its optimal solution, denoted as $\widetilde{\bf{R}} = \{\widetilde{R}^k_{m,n}| \forall k,m,n\}$. Based on the optimal solution to the continuous relaxation, we initialize the starting point for the steepest descent search as a feasible rate vector that is close to $\widetilde{\bf{R}}$, as $\mathbf{R}=\{\lfloor \widetilde{R}^k_{m,n}\rfloor_{\mathcal{R}} | \forall k,m,n\}$, where the operation $\lfloor \widetilde{R}^k_{m,n}\rfloor_{\mathcal{R}}= \arg \max_{R_u \in \mathcal{R}} R_u|R_u \le \widetilde{R}^k_{m,n}$ is defined such that $R^k_{m,n}$ takes the largest encoding rate value without exceeding $\widetilde{R}^k_{m,n}$ from the representation rate set $\mathcal{R}$. Starting from $\mathbf{R}$, we then search for the index $j^*$ that achieves the largest ratio of the distortion reduction to the rate increment for an increment in the rate element. If $j^*$ corresponds to the viewport tile of a user $k^*$, then according to the constraints in Eq. (\ref{eq:P1}f), the rates of all the viewport tiles of that user will be increased to the nearest higher rate value in $\mathcal{R}$. If $j^*$ corresponds to the marginal tile of a user $k^*$, then the rate of only that tile will be increased to the nearest higher rate value in $\mathcal{R}$. If the above rate increment does not violate the constraints in Eqs. (\ref{eq:P1}b)-(\ref{eq:P1}g), the optimal rate solution is updated accordingly; otherwise, the index $j^*$ is removed from the active searching set.

\section{Performance Evaluation}
\label{sec:simulate}

In this section, we evaluate the performance of the proposed steepest descent algorithm, and derive simple guidelines for the server-side rate adaptation for adaptive 360-degree video streaming under different simulation settings.

\subsection{Viewport Prediction}
We first evaluate the proposed CNN-based method on the dataset in \cite{Bao2016}. This dataset collected head motion data (represented by Euler angles) of 153 volunteers when they were watching 16 clips of 360-degree videos covering a variety of scenes, such as landscape and sport activities. Most of the volunteers only viewed a portion of the 16 clips, resulting in 985 views in total. Then through preprocessing, the head movement was sampled $10$ times per second, with each view containing $289$ samples. Therefore, the dataset includes a total of 285665 samples of the users' head motion. In the proposed CNN-based viewing angle prediction model, we use $80$ percent of the data for training, and the remaining $20$ percent for testing. To evaluate the accuracy of viewpoint prediction, we adopt three metrics, namely, mean error, root-mean-square error (RMSE), 99.9th percentile. For the implementation of the CNN-based model, the first convolutional layer has 32 channels, all the other convolutional layers have 64 channels, where Adam optimization \cite{Kingma2015} is used. The momentum and weight-decay are set to $0.8$ and $0.999$ respectively. We set the batch size to $128$ and train the model for $200$ epochs. The learning rate is linearly decayed from $1e-3$ to $1e-4$ in the first $100$ epochs.

\begin{table*}[!t]
\centering
\caption{Comparison on pitch and yaw angle prediction error of Naive, LR, NN and the proposed CNN-based model.}
\scalebox{0.715}{
\begin{tabular}
{|p{1.75cm}<\centering|p{1.5cm}<\centering|p{1.5cm}<\centering|p{1.25cm}<\centering|p{1.25cm}<\centering|p{1.25cm}<\centering|p{1.25cm}<\centering|p{1.25cm}<\centering|p{1.25cm}<\centering|p{1.25cm}<\centering|p{1.25cm}<\centering|p{1.25cm}<\centering|p{1.25cm}<\centering|}
\hline
  \multicolumn{3}{|c|}{Prediction time window (s)} & 0.1 & 0.2 & 0.3 & 0.4 & 0.5 & 0.6 & 0.7 & 0.8 & 0.9 & 1.0 \\ \hline
 \multirow{12}{*}{Pitch ${\theta\ (}^\circ)$} & \multirow{4}{*}{Mean}
 & Navie-x & 1.10 & 2.21 & 3.20 & 4.27 & 5.10 & 5.92 & 6.76 & 7.73 & 8.15 & 8.71 \\ \cline{3-13}
 & & LR-x & 1.14 & 2.81 & 5.26 & 7.13 & 8.89 & 12.37 & 13.84 & 16.09 & 18.52 & 21.51 \\ \cline{3-13}
 & & NN-x & \textbf{0.57} & 1.43 & 2.38 & 3.18 & 4.32 & 5.49 & 6.23 & 7.06 & 7.51 & 8.68 \\ \cline{3-13}
 & & CNN-x & 0.60 & \textbf{1.42} & \textbf{2.21} & \textbf{3.05} & \textbf{3.84} & \textbf{4.51} & \textbf{5.34} & \textbf{5.99} & \textbf{6.68} & \textbf{7.80} \\  \cline{2-13}
 & \multirow{4}{*}{RMSE}
 & Navie-x & 1.95 & 3.84 & 5.69 & 7.33 & 8.51 & 9.94 & 11.42 & 12.77 & 13.10 & 14.24 \\ \cline{3-13}
 & & LR-x & 1.85 & 4.10 & 6.70 & 8.75 & 10.60 & 14.43 & 16.18 & 18.40 & 21.11 & 24.26 \\ \cline{3-13}
 & & NN-x & \textbf{0.92} & 2.31 & 3.84 & 5.31 & 6.94 & 8.51 & 9.62 & 10.74 & 11.79 & 13.07 \\ \cline{3-13}
 & & CNN-x & 0.94 & \textbf{2.28} & \textbf{3.54} & \textbf{4.91} & \textbf{6.30} & \textbf{7.40} & \textbf{8.65} & \textbf{9.52} & \textbf{10.52} & \textbf{11.99} \\  \cline{2-13}
 & \multirow{4}{*}{99.9th}
 & Navie-x & 13.14 & 23.39 & 36.69 & 46.91 & 51.92 & 58.76 & 66.46 & 67.36 & 63.04 & 80.26 \\ \cline{3-13}
 & & LR-x & 10.73 & 25.98 & 37.25 & 41.73 & 43.74 & 57.36 & 51.38 & 65.39 & 70.14 & 74.96 \\ \cline{3-13}
 & & NN-x & 6.75 & 17.45 & 27.66 & 37.02 & 46.13 & 52.28 & 57.19 & 60.99 & 69.04 & 72.02 \\ \cline{3-13}
 & & CNN-x & \textbf{6.62} & \textbf{14.61} & \textbf{22.90} & \textbf{31.05} & \textbf{41.22} & \textbf{45.41} & \textbf{54.61} & \textbf{54.86} & \textbf{60.42} & \textbf{69.24} \\ \hline
 \multirow{12}{*}{Yaw ${\varphi\ (}^\circ)$} & \multirow{4}{*}{Mean}
 & Navie-y & 2.68 & 5.40 & 7.86 & 10.21 & 12.03 & 14.93 & 16.94 & 18.76 & 20.66 & 22.91 \\ \cline{3-13}
 & & LR-y & 2.68 & 5.50 & 8.43 & 10.21 & 13.14 & 16.16 & 18.71 & 20.90 & 22.27 & 25.57 \\ \cline{3-13}
 & & NN-y & 0.91 & 2.41 & 4.23 & 6.20 & 8.23 & 10.18 & 12.29 & 13.99 & 15.93 & 17.78 \\ \cline{3-13}
 & & CNN-y & \textbf{0.88} & \textbf{2.29} & \textbf{4.08} & \textbf{5.88} & \textbf{7.75} & \textbf{9.49} & \textbf{11.47} & \textbf{13.24} & \textbf{14.90} & \textbf{16.55} \\ \cline{2-13}
 &\multirow{4}{*}{RMSE}
 & Navie-y & 4.87 & 9.73 & 13.72 & 17.45 & 20.25 & 24.72 & 27.67 & 29.95 & 32.64 & 36.16 \\ \cline{3-13}
 & & LR-y & 4.58 & 9.77 & 14.47 & 16.96 & 20.41 & 24.42 & 28.78 & 30.65 & 32.48 & 36.86 \\ \cline{3-13}
 & & NN-y & 1.83 & 4.59 & 7.82 & 11.45 & 14.60 & 17.55 & 20.89 & 23.25 & 26.06 & 28.46 \\ \cline{3-13}
 & & CNN-y & \textbf{1.57} & \textbf{3.90} & \textbf{7.13} & \textbf{10.02} & \textbf{13.04} & \textbf{15.82} & \textbf{18.71} & \textbf{21.58} & \textbf{24.12} & \textbf{26.31} \\ \cline{2-13}
 &\multirow{4}{*}{99.9th}
 & Navie-y & 32.58 & 70.31 & 91.17 & 114.51 & 123.99 & 143.91 & 158.62 & 165.69 & 163.54 & 170.73 \\ \cline{3-13}
 & & LR-y & 27.04 & 64.26 & 95.82 & 125.25 & 137.65 & 116.71 & 158.55 & 157.09 & 161.84 & 158.57 \\ \cline{3-13}
 & & NN-y & 14.07 & 37.57 & 68.47 & 92.45 & 112.75 & 135.87 & 152.55 & 157.66 & 160.58 & 165.56 \\ \cline{3-13}
 & & CNN-y & \textbf{11.87} & \textbf{31.65} & \textbf{53.28} & \textbf{71.70} & \textbf{94.63} & \textbf{115.92} & \textbf{121.69} & \textbf{137.89} & \textbf{157.86} & \textbf{155.26} \\ \hline
\end{tabular}}\vspace*{-0.6cm}
\end{table*}

In terms of these three metrics, we compare the prediction accuracy of the proposed CNN-based viewpoint prediction model with three existing prediction models, naive \cite{Qian2016}, linear regression (LR) \cite{Bao2016} and neural network (NN) \cite{Bao2016}. As shown in Table I, the proposed CNN-based model significantly improves the prediction accuracy for both the pitch and yaw angles, especially when the prediction window becomes large. Compared to the NN model that performs best among the prior works, the proposed CNN-based model increases the prediction accuracy by $9\% \sim 27\%$ in all the three metrics, and achieves up to $69\%$ improvement in comparison to the baseline naive model. This indicates that the proposed CNN-based model has a stronger ability of nonlinearity fitting and performs better for a larger prediction window.

\subsection{Tile Rate Allocation}
To evaluate the performance of the proposed server-side tile rate adaptation optimization and the steepest descent algorithm, we further compare the performance of the proposed steepest descent algorithm in Algorithm 1 with two other tile-based rate adaptation schemes: \textit{baseline}, where all the tiles are transmitted to the users at the same quality (encoding rate); and \textit{greedy}, where tiles in the predicted viewport region of users are allocated with a high quality (encoding rate) while the remaining tiles are allocated with the lowest quality (encoding rate). We use three recommended test sequences for 360-degree videos, \textit{Canolafield} \cite{Drive2016}, \textit{Driving in Country} \cite{Drive2016} and \textit{Basketball} \cite{Basketball2016}, with an identical spatial resolution of $3840 \times 1920$ and the same frame rate of $25$ fps. We divide the 360-degree video in ERP format into $8 \times 8$ tiles and use high efficiency video coding (HEVC) to encode each tile into six different rate representations. The corresponding QP for encoding each rate representation is set as $\{42, 37, 32, 27, 22, 17\}$, respectively, resulting in an increasing average encoding rate per tile as $\{R_1, R_2, R_3, R_4, R_5, R_6\}=\{2, 4, 10, 20, 49, 120\}$ kbps. For each test sequence, ten users are simultaneously watching that video wearing the same type of HMD, i.e., the perspective FoVs of their HMDs are $90^\circ$ vertically and $110^\circ$ horizontally. However, these users will watch the same video with different initial viewpoints and head movement trajectories, which are set accordingly based on ten randomly selected views from the head motion dataset in \cite{Bao2016}. We use the prediction time window of 1.0 s and set the probability threshold to differentiate the marginal and invisible tiles for each user as $\alpha=0.05$.

\begin{figure}[!t]
    \centering
     \subfigure[]{ \label{fig_a}
    \includegraphics[width=4.25in]{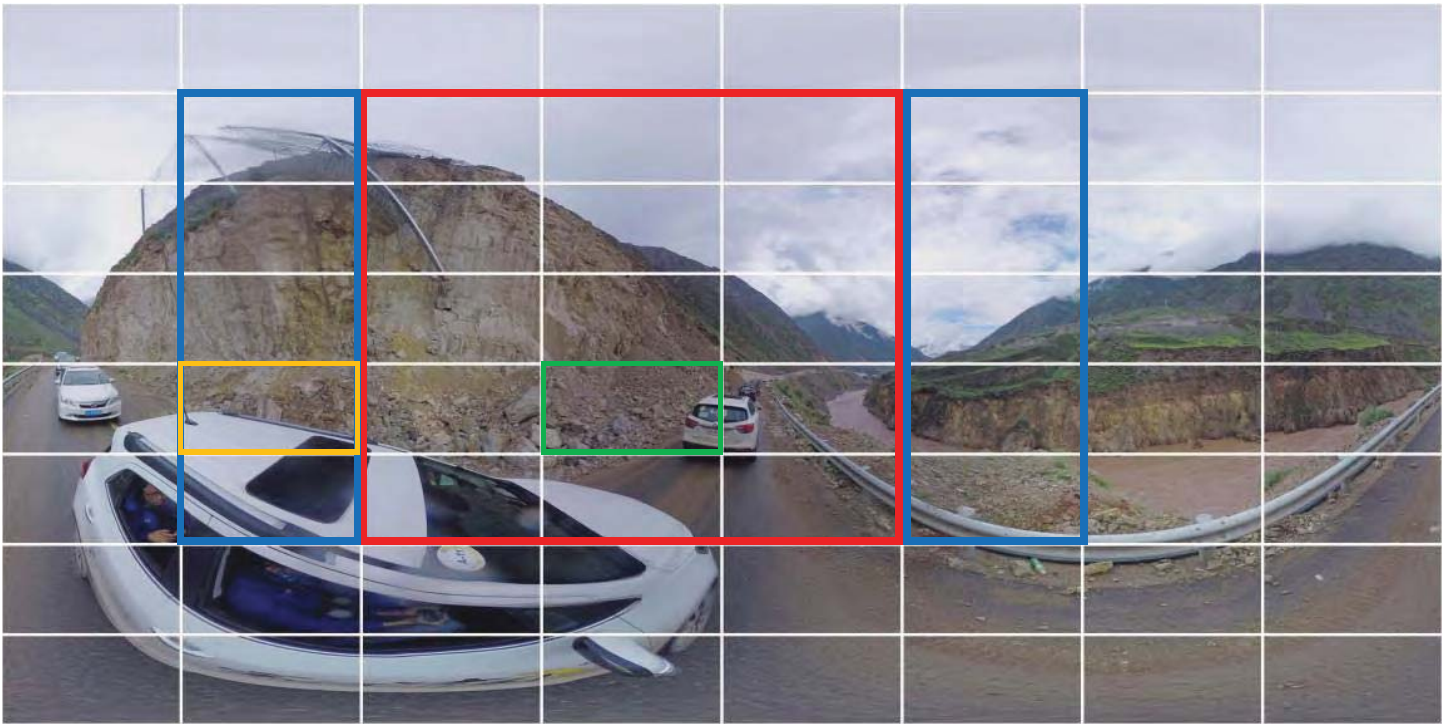} }
    \subfigure[]{ \label{fig_b}
    \includegraphics[width=4.25in]{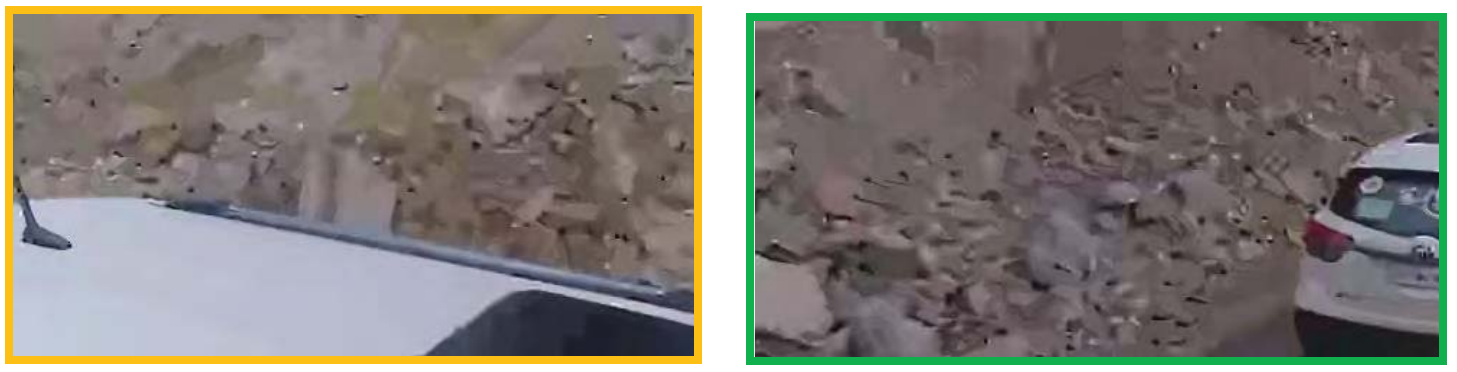} }
     \subfigure[]{ \label{fig_c}
    \includegraphics[width=4.25in]{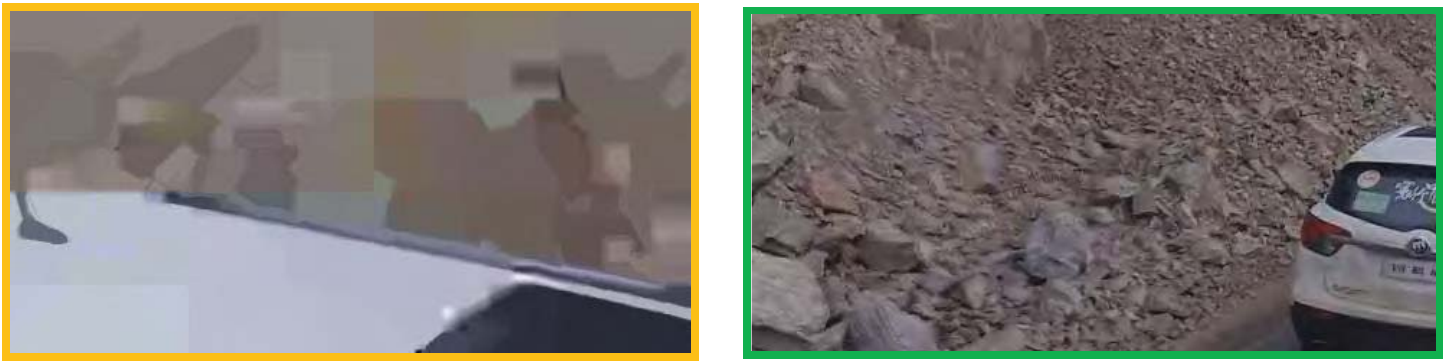} }
    \subfigure[]{ \label{fig_d}
    \includegraphics[width=4.25in]{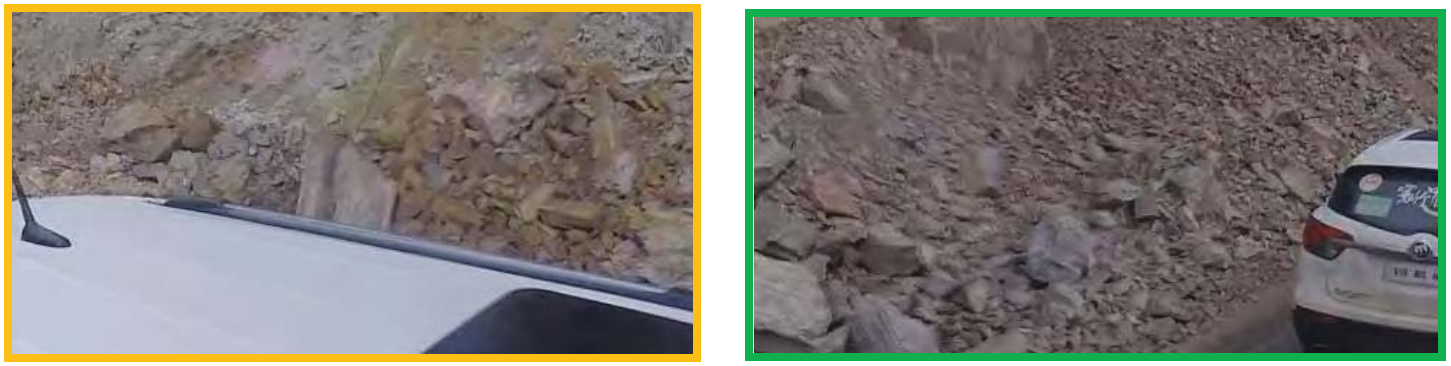} }
    \caption{(a) Tiling in a frame of \textit{Driving in Country}, where tiles inside the red rectangle represent predicted viewport tiles for a user and tiles inside the blue rectangles represent marginal tiles. Illustration of the achievable subjective quality for two tiles with (b) baseline algorithm, (c) greedy algorithm, and (d) proposed algorithm.}
    \label{fig:Drive}\vspace*{-0.7cm}
\end{figure}

Fig. \ref{fig:Drive}(a) illustrates a tiling example for a user in a frame of the video \textit{Driving in Country}, where tiles in the red rectangle represent the predicted viewport tiles and tiles in the blue rectangles represent marginal tiles. Then, in Figs. \ref{fig:Drive}(b)-\ref{fig:Drive}(d), we compare the quality of a viewport tile (green rectangle) and a marginal tile (yellow rectangle) achieved by the three different algorithms. Here, the server's transmission capacity is set to $C^s=26$ Mbps, while all the ten users are assumed to have the same transmission capacity $C^k=2$ Mbps. In this case, the user's transmission capacity is not sufficient to guarantee the transmission of visible tiles with the highest quality. Since the \textit{baseline} algorithm allocates the user's transmission capacity equally to all the tiles within the frame, both the viewport tile and the marginal tile are transmitted with a relatively low bitrate representation. It can be seen in Fig. \ref{fig:Drive}(b) that the visual quality of both the viewport tile and the marginal tile presented to the user is the same but at a relatively low level. The \textit{greedy} algorithm attempts to allocate a highest possible bitrate representation for the tiles within the predicted viewport region subject to the transmission capacity, while the marginal and invisible tiles are allocated with the lowest bitrate representation. The user's transmission capacity of 2 Mbps allows the transmission of viewport tiles with the highest bitrate representation, however, the marginal tiles are still delivered with the lowest bitrate representation. It, as shown in Fig. \ref{fig:Drive}(c), results in an uncomfortable quality degradation when the predicted user's viewport deviates from the actual viewport such that some marginal tiles are included in the actual viewport. Compared with these two algorithms, as shown in Fig. \ref{fig:Drive}(d), the proposed algorithm not only allocates a relatively high bitrate representation for the predicted viewport tiles, but also preserves for the marginal tiles with a representation the bitrate of which is lower than the viewport tile but still higher than the lowest bitrate representation to tolerate the viewport prediction error.

\begin{figure}[!t]
    \centering
    \subfigure[]{ \label{fig_a}
    \includegraphics[width=2.625in]{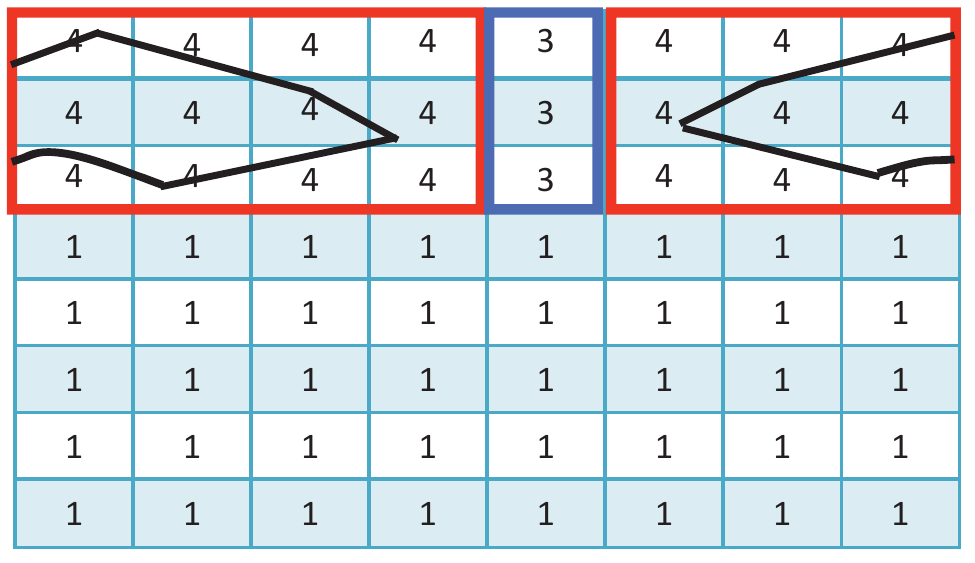} }
     \subfigure[]{ \label{fig_b}
    \includegraphics[width=2.7in]{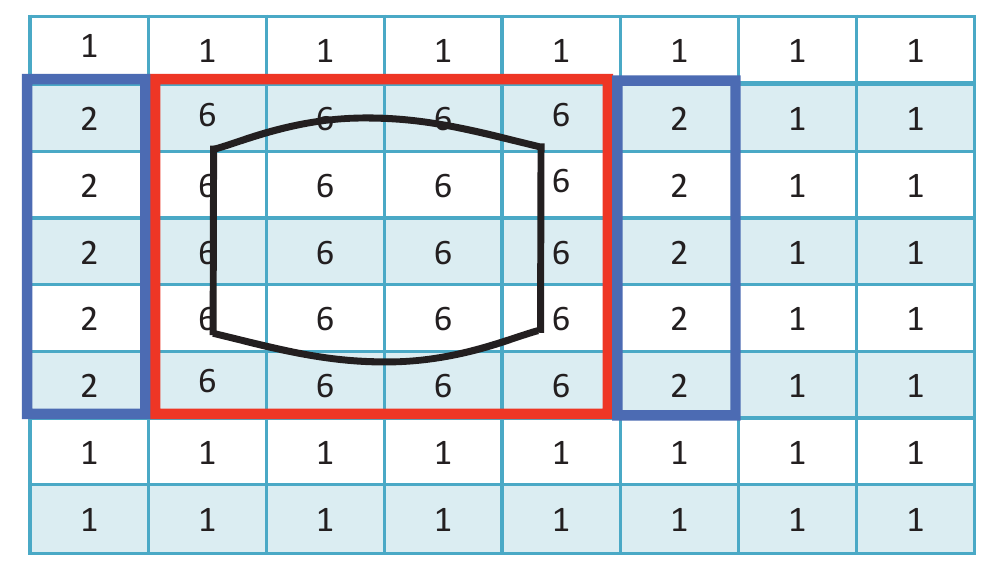} }
    \caption{Representation rate indices of tiles received by two users when using the proposed algorithm, where $\omega=0$, $C^s=26$ Mbps and $C^k=2$ Mbps: (a) user's viewpoint is near the north pole, and (b) user's viewpoint is near the equator. (Black lines cover the predicted viewport region while the red rectangles cover the viewport tiles, and blue rectangles represent the marginal tiles.)}
    \label{fig:DifferUserW0}\vspace*{-0.2cm}
\end{figure}

In Fig. \ref{fig:DifferUserW0}, we compare the representation rate indices of tiles allocated by the proposed algorithm for two users with different viewports, where $\omega=0$, $C^s=26$ Mbps and $C^k=2$ Mbps. Comparing the location of these two viewports, the user's viewport in Fig. \ref{fig:DifferUserW0}(a) is close to the north pole while that in Fig. \ref{fig:DifferUserW0}(b) is near the equator. Therefore, a higher rate representation (i.e., $R_6$) is allocated for viewport tiles in Fig. \ref{fig:DifferUserW0}(b) than the rate representation (i.e., $R_4$) allocated for those in Fig. \ref{fig:DifferUserW0}(a). This is because we take the WS-PSNR measure to define the weighted average distortion of tiles as the 360-degree video distortion. The weight equals to 1 when the tile is around the equator and decreases towards zero when it becomes closer to the poles. Therefore, the viewport tiles in Fig. \ref{fig:DifferUserW0}(b) have higher weights than those in Fig. \ref{fig:DifferUserW0}(a), and thus are allocated with a higher rate representation. On the other hand, the marginal tiles in Fig. \ref{fig:DifferUserW0}(a) are allocated with a higher rate representation (i.e., $R_3$) than the rate representation (i.e., $R_2$) of those in Fig. \ref{fig:DifferUserW0}(b), since according to Eq. (\ref{eq:Pvisible}) their visible probabilities are higher.

\begin{figure}[!t]
    \centering
    \subfigure[]{ \label{fig:fa}
    \includegraphics[width=2.7in]{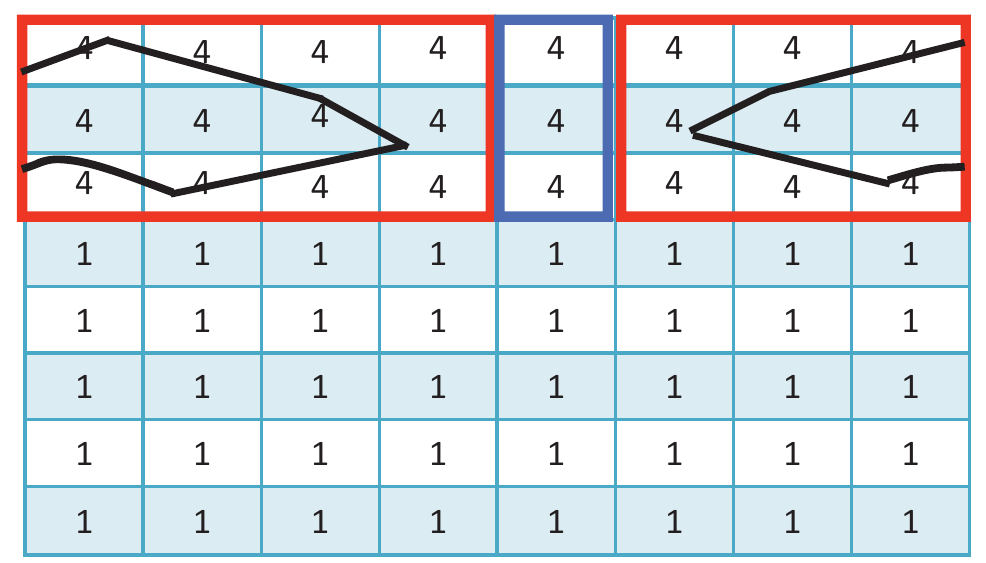} }
     \subfigure[]{ \label{fig:fb}
    \includegraphics[width=2.7in]{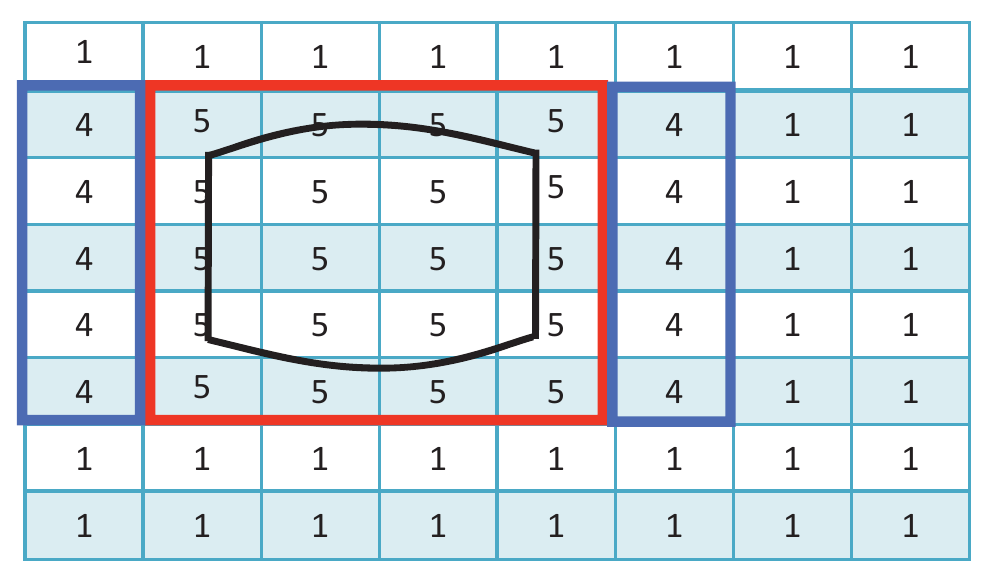} }
    \caption{Representation rate indices of tiles received by two users when using the proposed algorithm, where $\omega=1$, $C^s=26$ Mbps and $C^k=2$ Mbps: (a) user's viewpoint is near the north pole, and (b) user's viewpoint is near the equator. (Black lines cover the predicted viewport region while the red rectangles cover the viewport tiles, and blue rectangles represent the marginal tiles.)}
    \label{fig:DifferUserW1}\vspace*{-0.7cm}
\end{figure}

In Fig. \ref{fig:DifferUserW1}, we increase the value of weight parameter $\omega$ to one, and keep the other experiment settings the same as in Fig. \ref{fig:DifferUserW0}. Through comparison, it can be seen that with the increment of the weight parameter $\omega$, the difference in representation rate (or quality) between the viewport tiles and marginal tiles of the same user is reduced. This is because according to the optimization objective in Eq. (\ref{eq:P1}a), a larger $\omega$ will impose a higher penalty on the representation rate difference between the viewport and marginal tiles. In this way, the quality difference between these two types of tiles is flattened, and the quality degradation when the user switches the viewing direction from the predicted viewport to the marginal region becomes imperceptible.

\begin{figure*}[!t]
\centering
    \subfigure[]{ \label{fig_a}
    \includegraphics[width=3.05in]{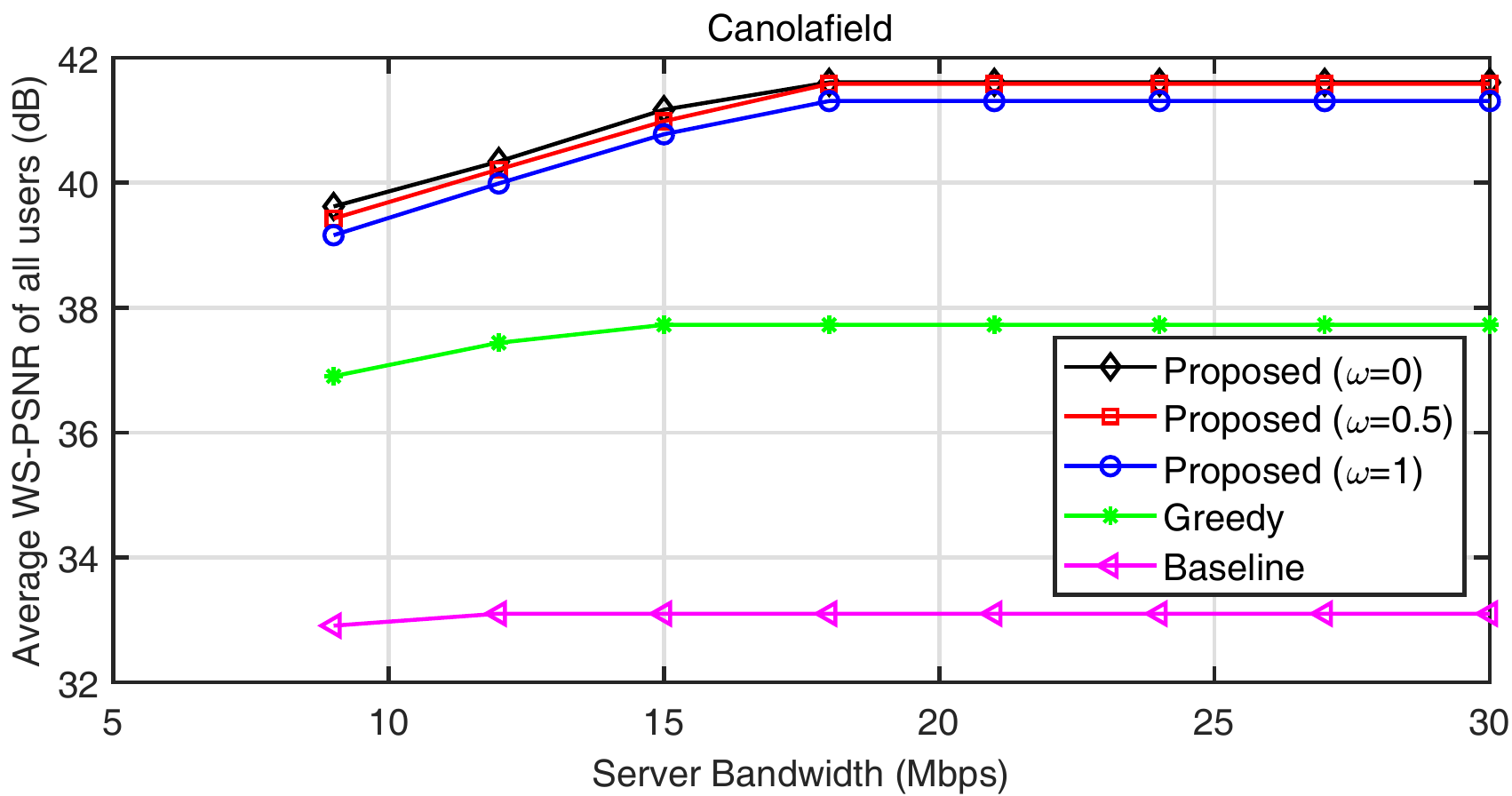} }
    \subfigure[]{ \label{fig_b}
    \includegraphics[width=3.05in]{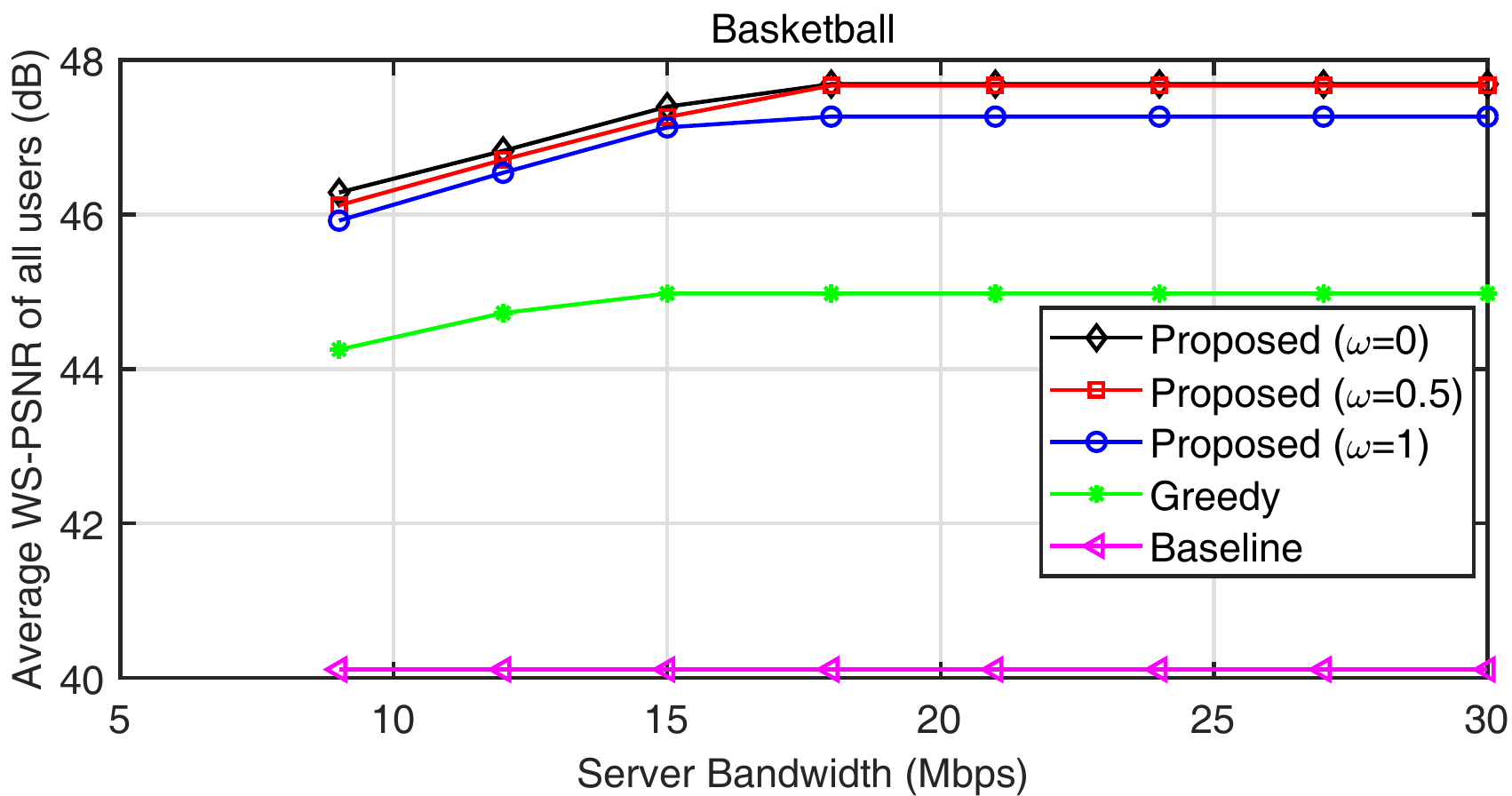} }
    \subfigure[]{ \label{fig_b}
    \includegraphics[width=3.05in]{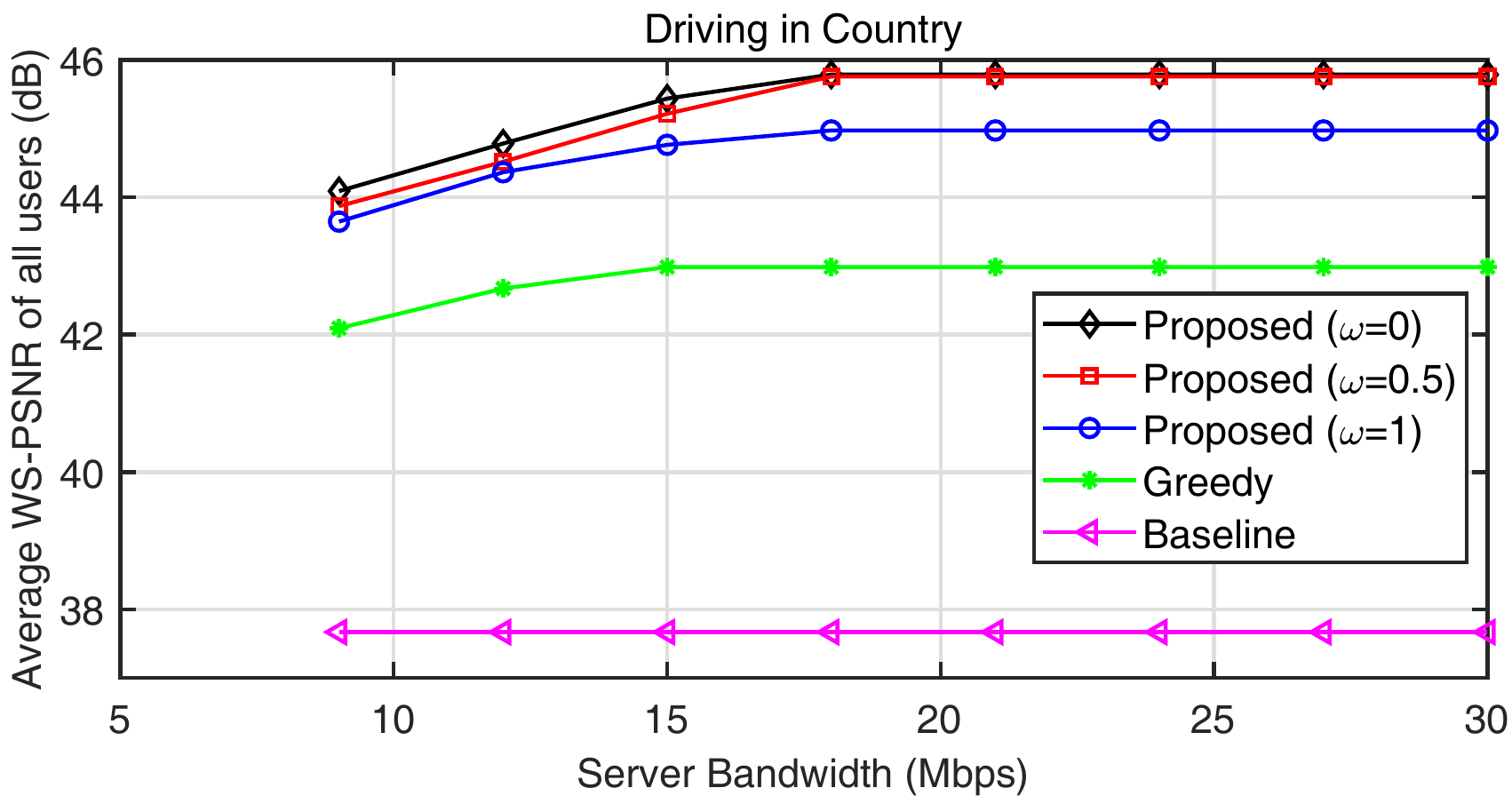} }
    \subfigure[]{ \label{fig_b}
    \includegraphics[width=3.05in]{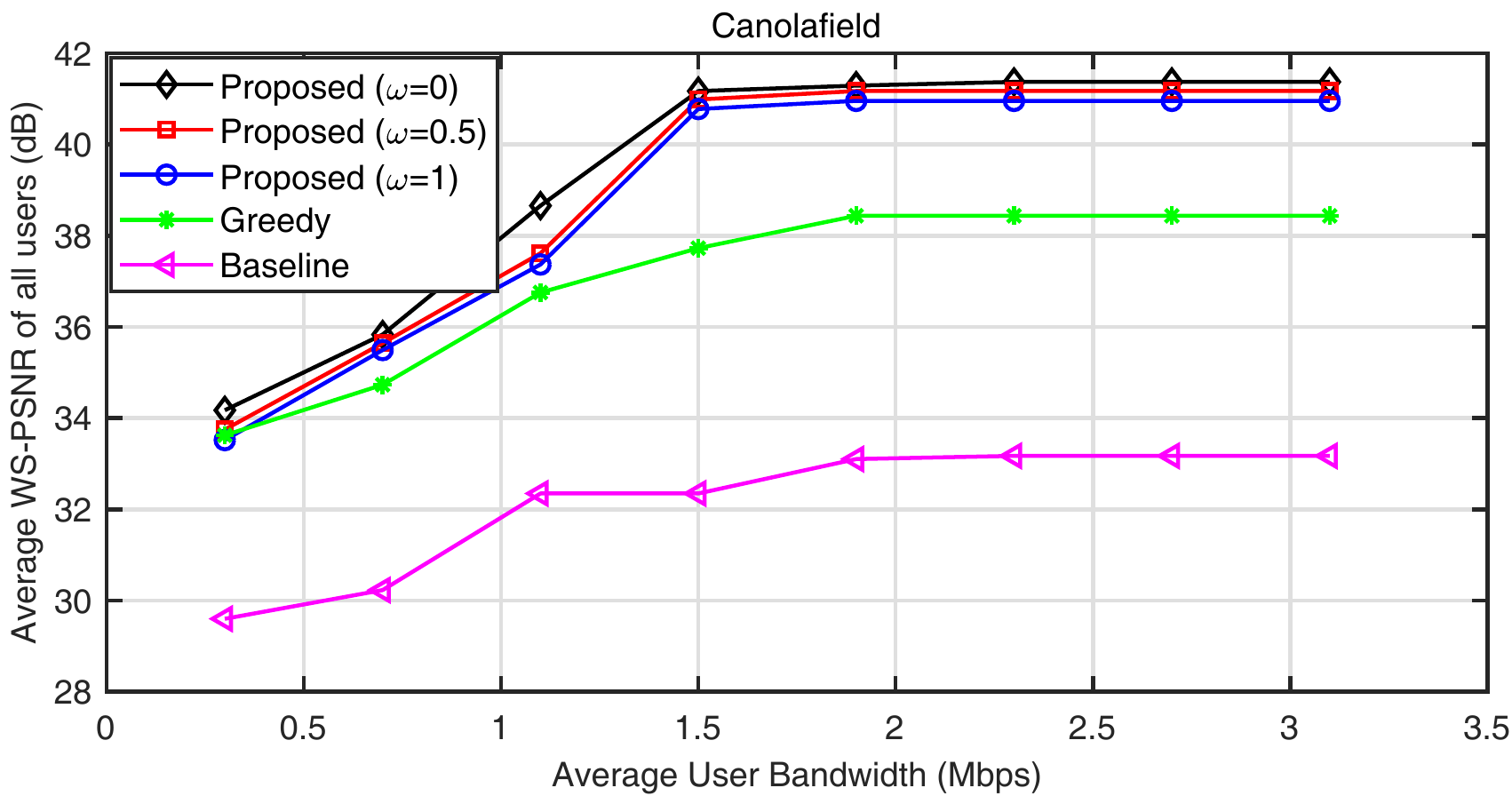} }
    \subfigure[]{ \label{fig_b}
    \includegraphics[width=3.05in]{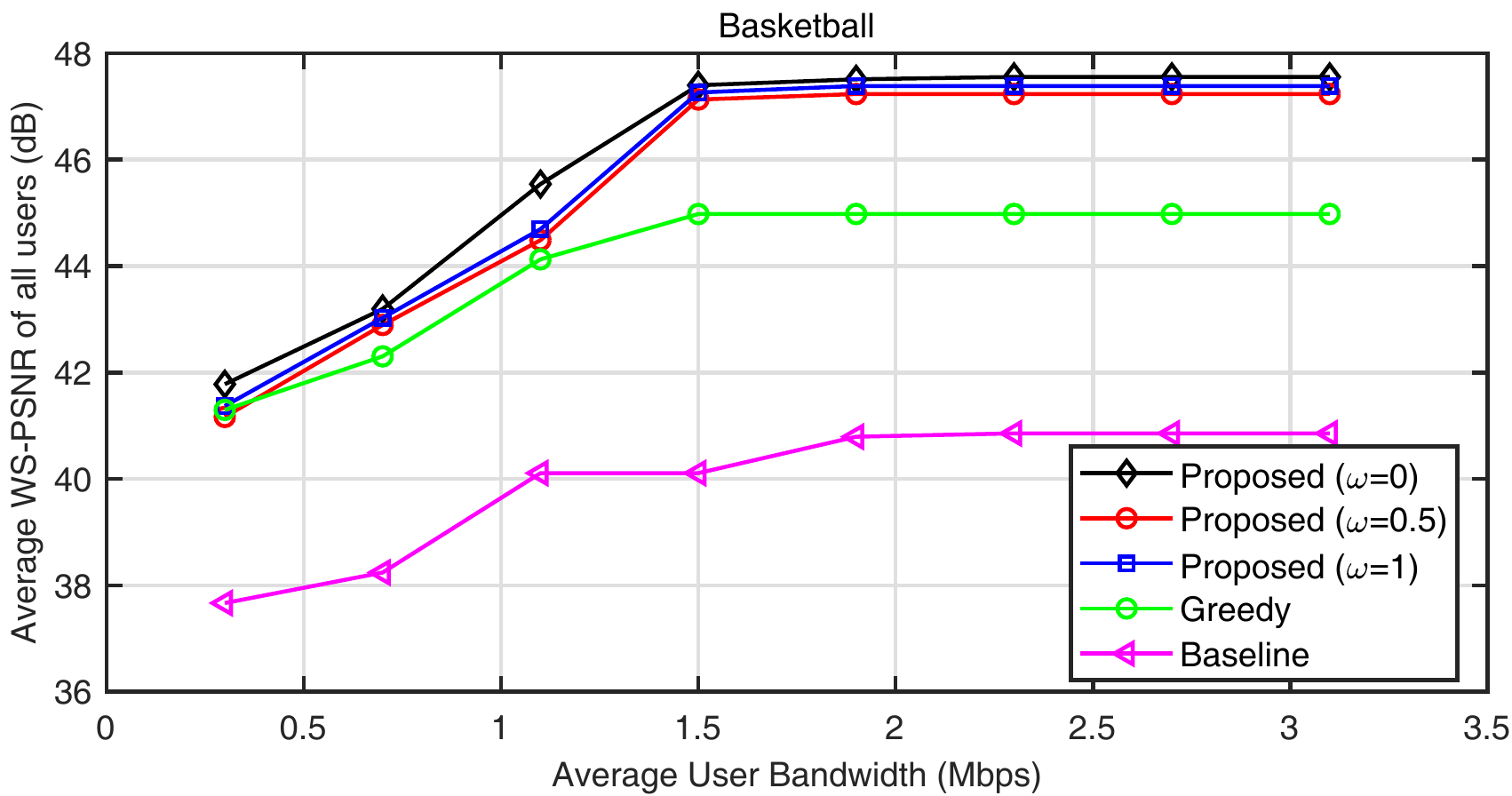} }
    \subfigure[]{ \label{fig_b}
    \includegraphics[width=3.05in]{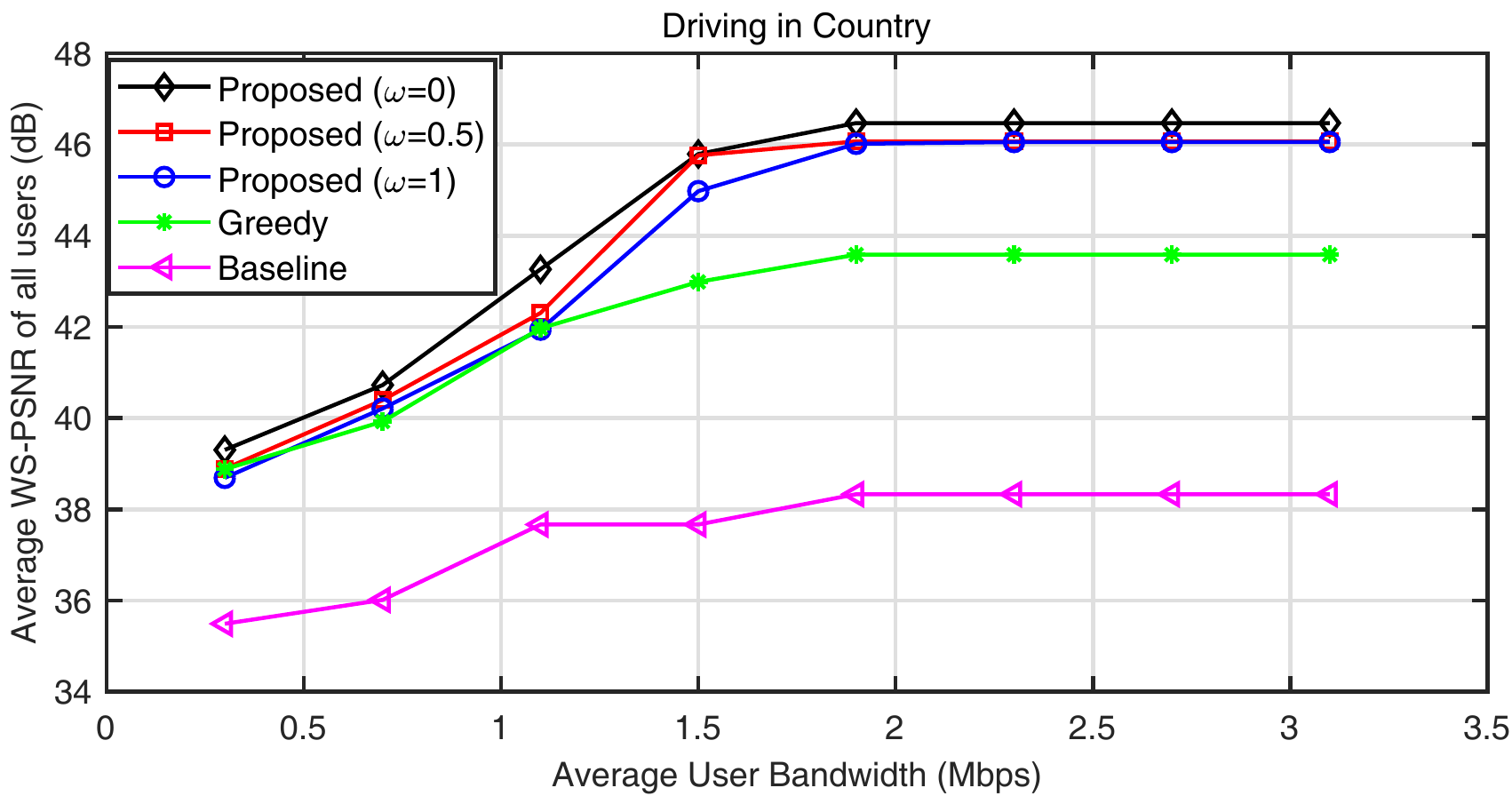} }
    \caption{Average WS-PSNR of all users vs. server's transmission capacity for (a) \textit{Canolafield}, (b) \textit{Basketball}, and (c) \textit{Driving in Country}; and average WS-PSNR of all users vs. average user transmission capacity for (d) \textit{Canolafield}, (e) \textit{Basketball}, and (f) \textit{Driving in Country}.}
    \label{fig:PSNR}\vspace*{-0.7cm}
\end{figure*}

Fig. \ref{fig:PSNR} compares the average WS-PSNR of all users when they are viewing the three test video sequences with different algorithms, under different settings of server's and users' transmission capacities. Specifically, in Figs. \ref{fig:PSNR}(a)-\ref{fig:PSNR}(c), we assume that the transmission capacity of each user is a randomly selected value following the uniform distribution within the range of $[1.3,1.7]$ Mbps, and vary the server's transmission capacity from 9 Mbps to 30 Mbps. It can be seen that the average WS-PSNR versus the server's transmission capacity curves achieved by the same algorithm for the three sequences present the same trend. For the same video sequence, the average WS-PSNR of all the three algorithms will increase as the server's transmission capacity becomes larger. When the server's transmission capacity exceeds the sum of transmission capacities of all the users, the average WS-PSNR achieved by each algorithm will stop increasing. The proposed algorithm generally outperforms the \textit{baseline} and \textit{greedy} algorithms by achieving a much higher average WS-PSNR for a given transmission capacity of the server. In addition, as already discussed in Figs. \ref{fig:DifferUserW0} and \ref{fig:DifferUserW1}, the average WS-PSNR performance will decrease slightly with the increment of weight parameter $\omega$, but the quality degradation between the viewport and marginal tiles will become imperceptible. In Figs. \ref{fig:PSNR}(d)-\ref{fig:PSNR}(f), we set the server's transmission capacity as 15 Mbps, and vary the average user transmission capacity from 0.3 Mbps to 3.1 Mbps, where similar observations can be found for the average WS-PSNR versus average user transmission capacity curves achieved by different algorithms.

\begin{figure*}[!t]
\centering
    \subfigure[]{ \label{fig_a}
    \includegraphics[width=3.05in]{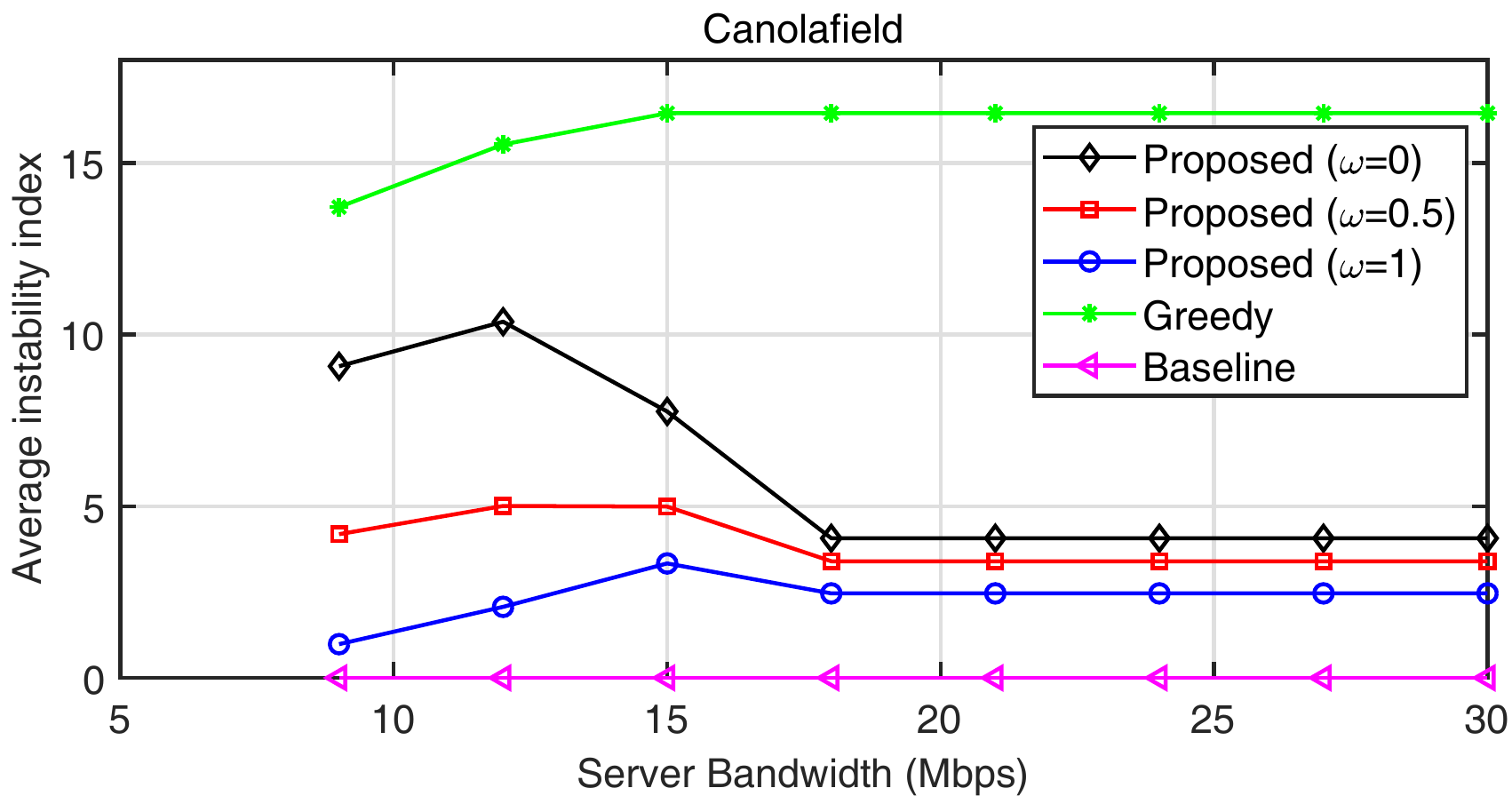} }
    \subfigure[]{ \label{fig_b}
    \includegraphics[width=3.05in]{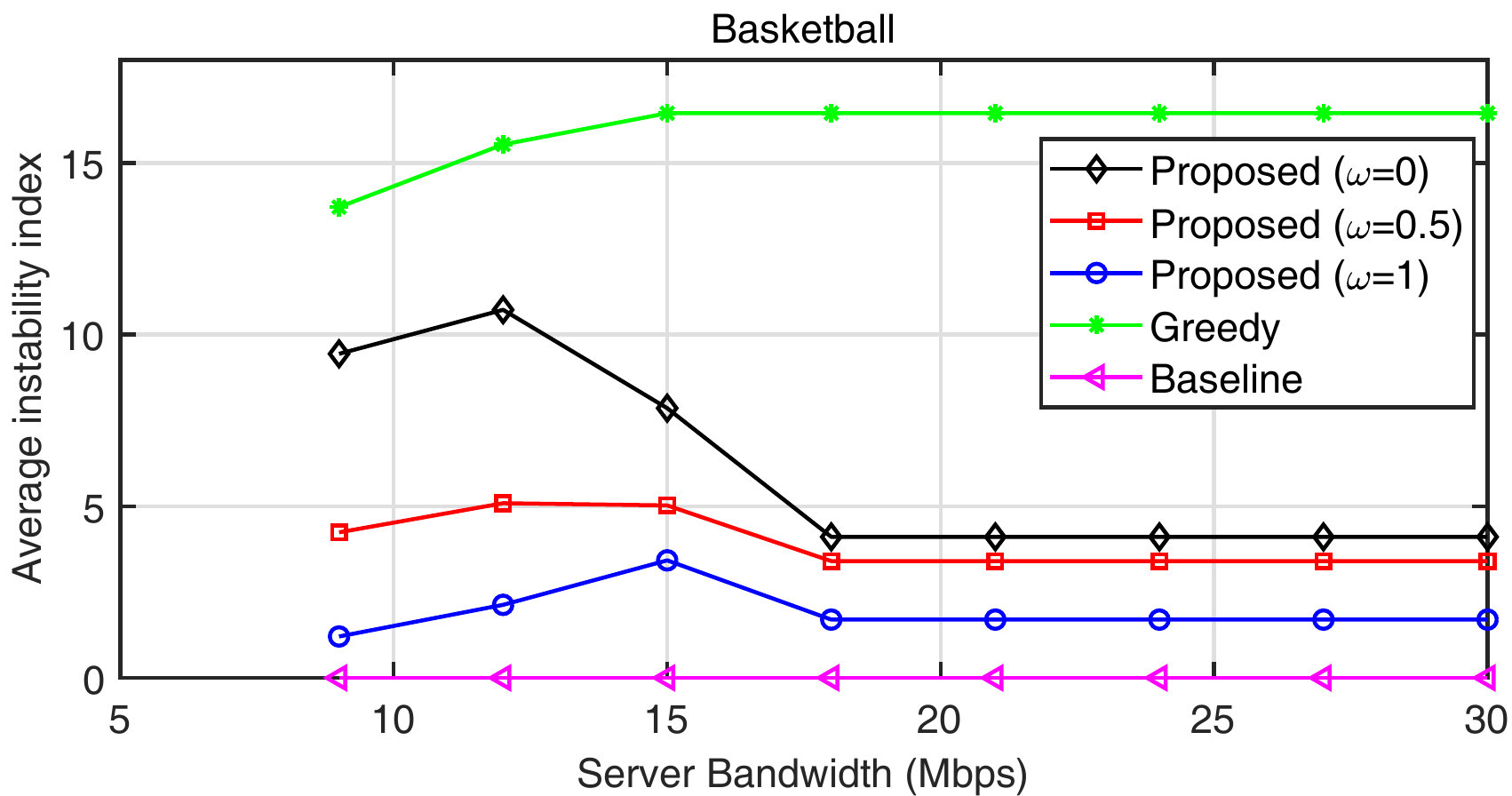} }
    \subfigure[]{ \label{fig_b}
    \includegraphics[width=3.05in]{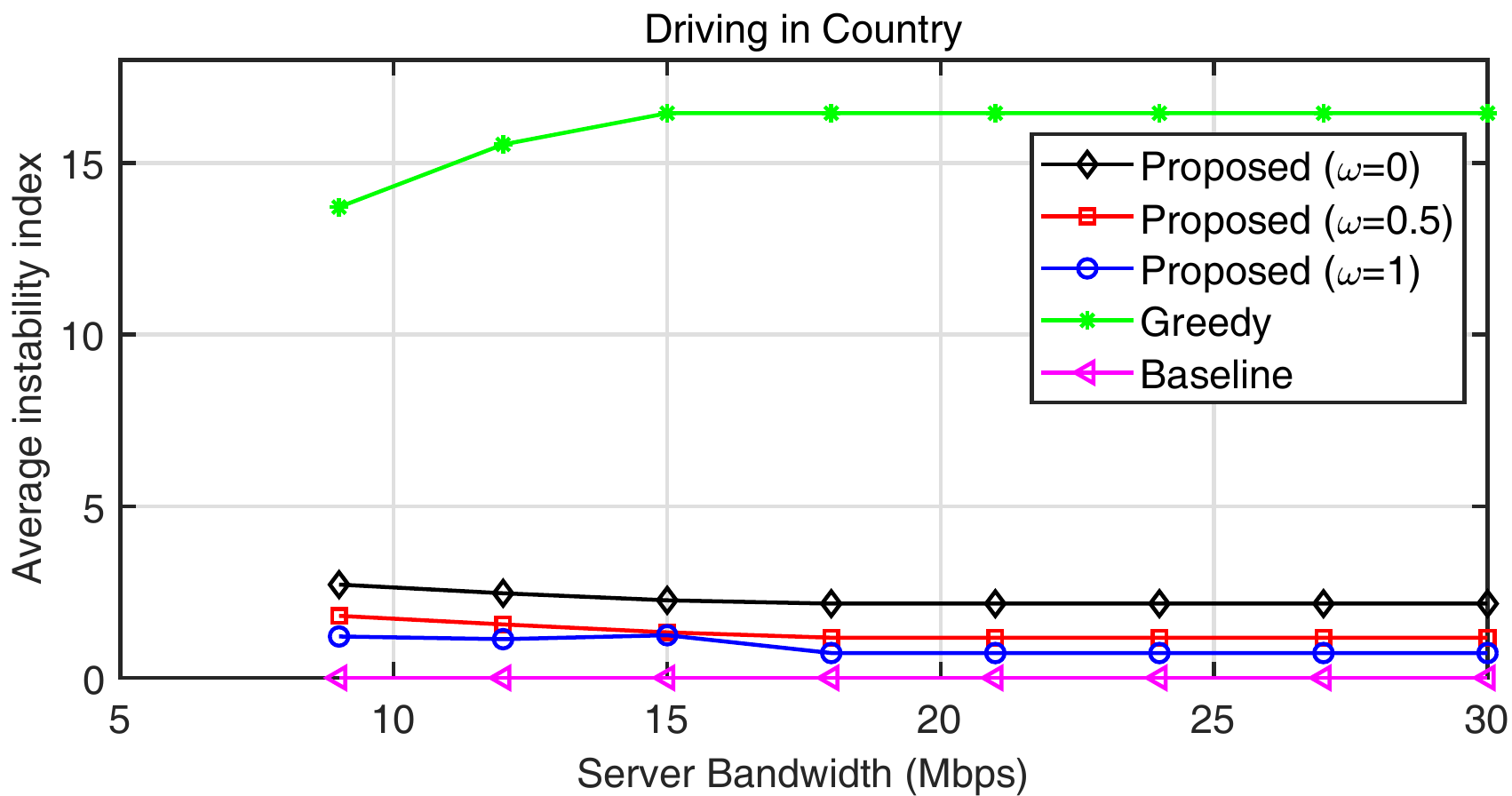} }
    \subfigure[]{ \label{fig_b}
    \includegraphics[width=3.05in]{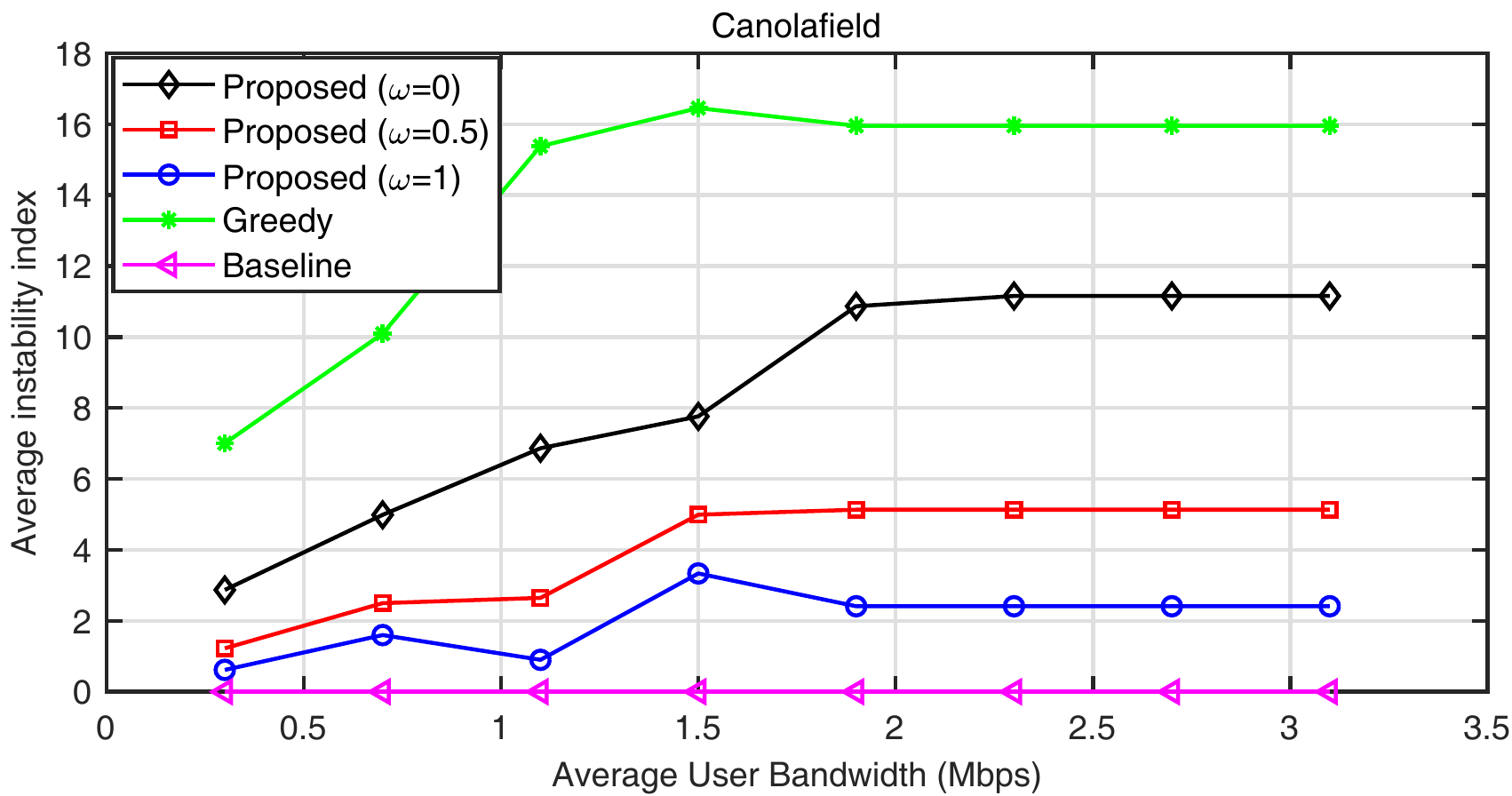} }
    \subfigure[]{ \label{fig_b}
    \includegraphics[width=3.05in]{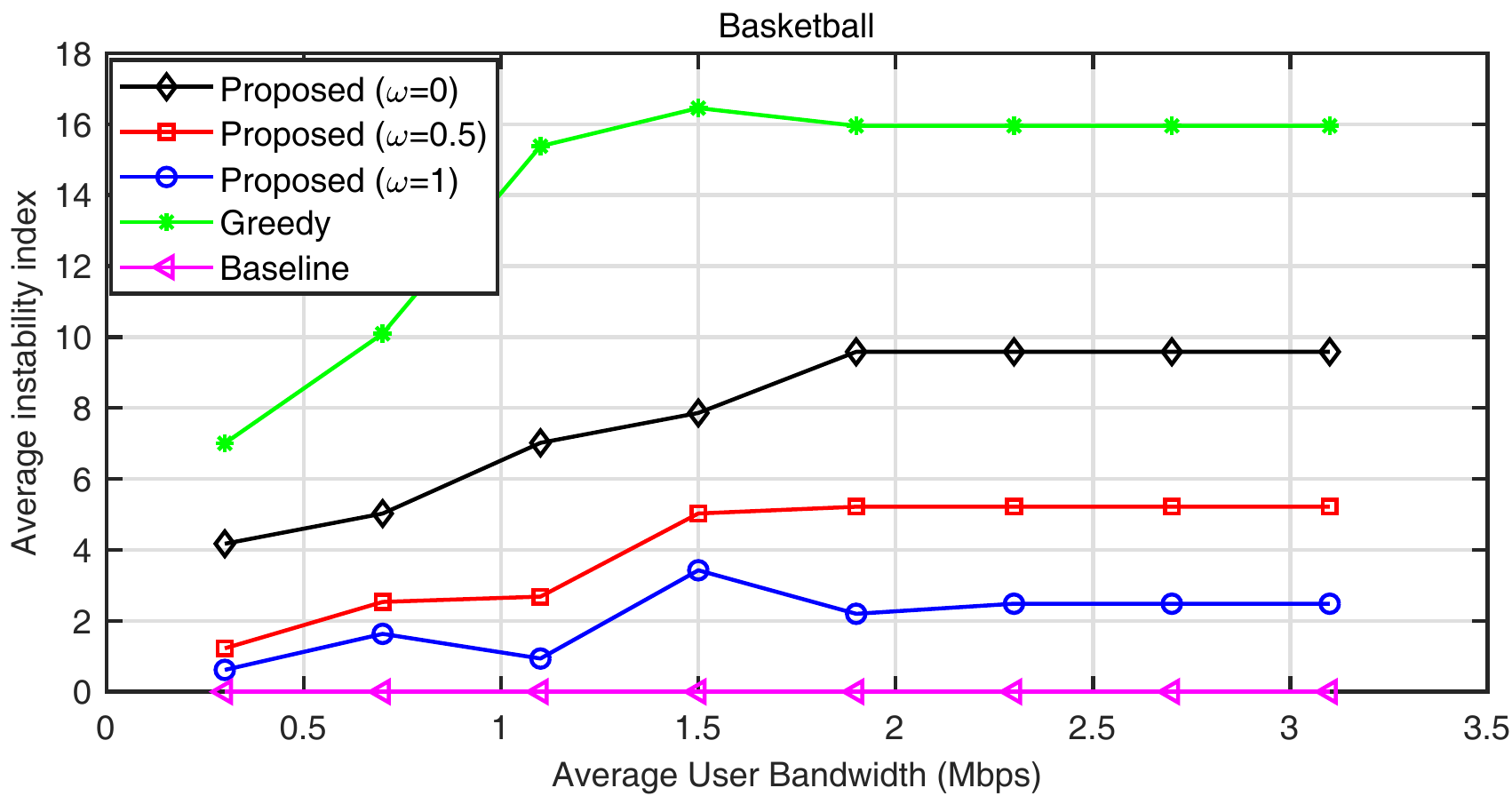} }
    \subfigure[]{ \label{fig_b}
    \includegraphics[width=3.05in]{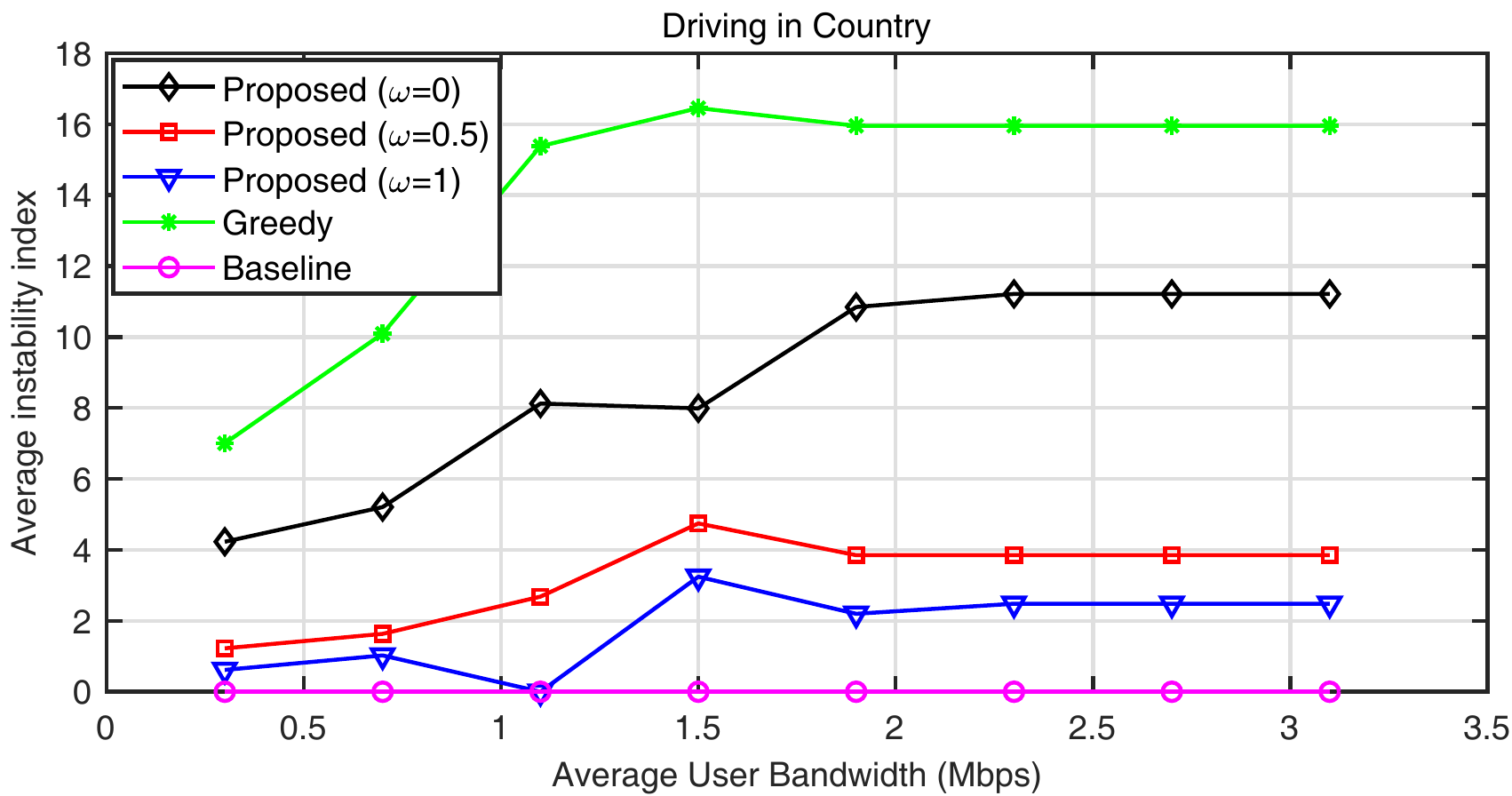} }
\caption{Average instability index of all users vs. server's transmission capacity for (a) \textit{Canolafield}, (b) \textit{Basketball}, and (c) \textit{Driving in Country}; and average instability index of all users vs. average user transmission capacity for (d) \textit{Canolafield}, (e) \textit{Basketball}, and (f) \textit{Driving in Country}.}
\label{fig:sta}\vspace*{-0.7cm}
\end{figure*}

To reflect the quality difference between viewport and marginal tiles, we define an instability index for each user as the standard deviation of the representation rate indices of all the tiles within the viewport and marginal regions. Further, in Fig. \ref{fig:sta}, we compare the average instability index of all the users achieved by different algorithms. Unless stated otherwise in the figure, the server's transmission capacity is set to 15 Mbps, while the transmission capacity of each user is a randomly selected value that follows the uniform distribution within the range of $[1.3,1.7]$ Mbps. It can be observed that the instability index of the \textit{baseline} algorithm is always zero, while the \textit{greedy} algorithm presents the largest instability index under the same transmission capacity setting for the server and users. This is because all the tiles within the frame are allocated with the same rate representation by the \textit{baseline algorithm}, and the \textit{greedy} algorithm attempts to allocate a highest possible bitrate representation for viewport tiles while the other tiles are allocated with the lowest bitrate representation. In comparison to the \textit{greedy} algorithm, the proposed algorithm preserves for the marginal tiles with a lower representation rate than the viewport tile but still higher than the lowest bitrate representation. Therefore, the proposed algorithm achieves a much smaller instability index than the \textit{greedy} algorithm, for given transmission capacities of the server and users. In addition, when the weight parameter $\omega$ increases from zero to one, the instability index can be further reduced to become more tolerant to the viewport prediction error.

\begin{figure*}[!t]
\centering
    \subfigure[]{ \label{fig_a}
    \includegraphics[width=3.05in]{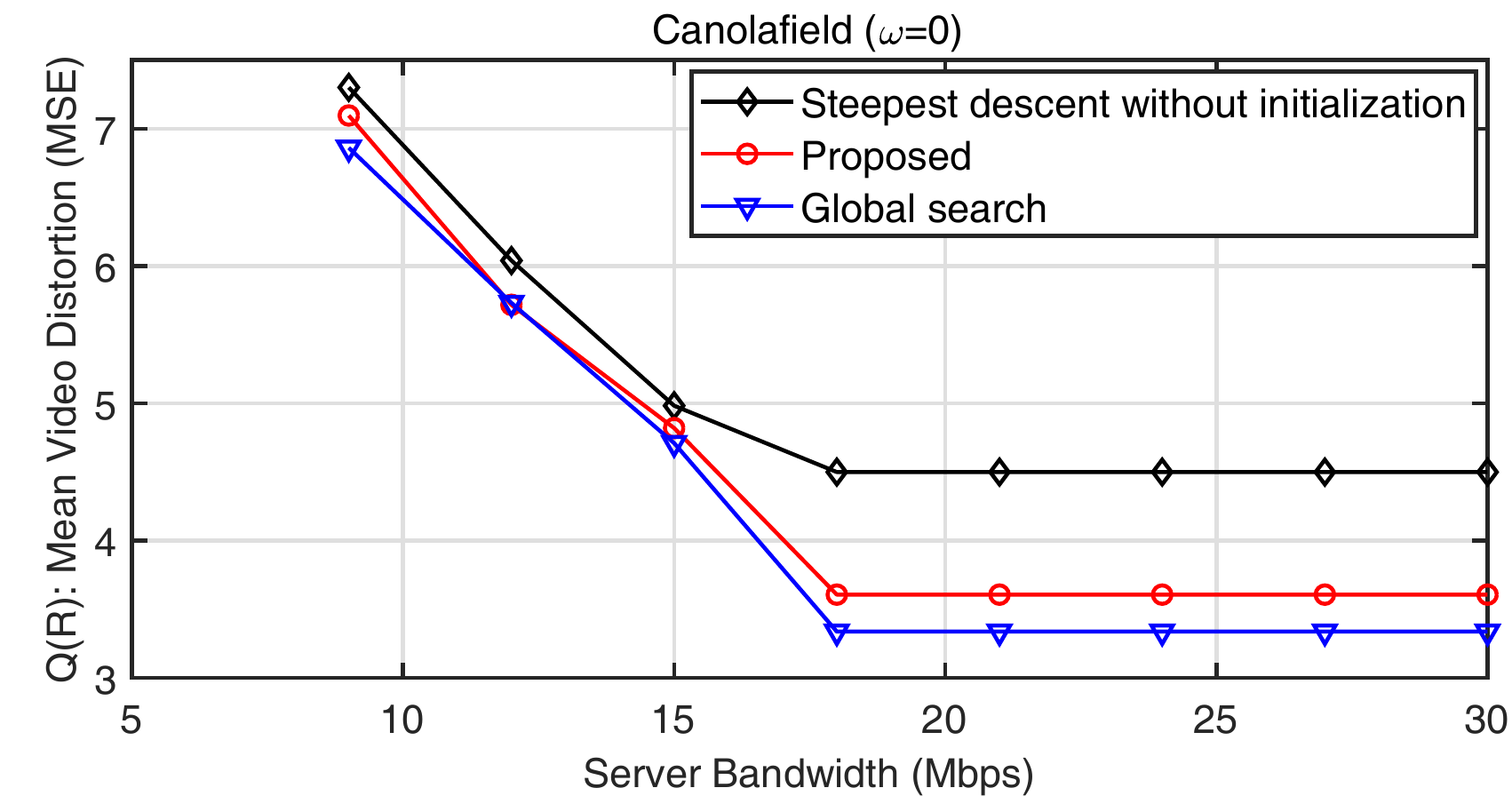} }
    \subfigure[]{ \label{fig_b}
    \includegraphics[width=3.05in]{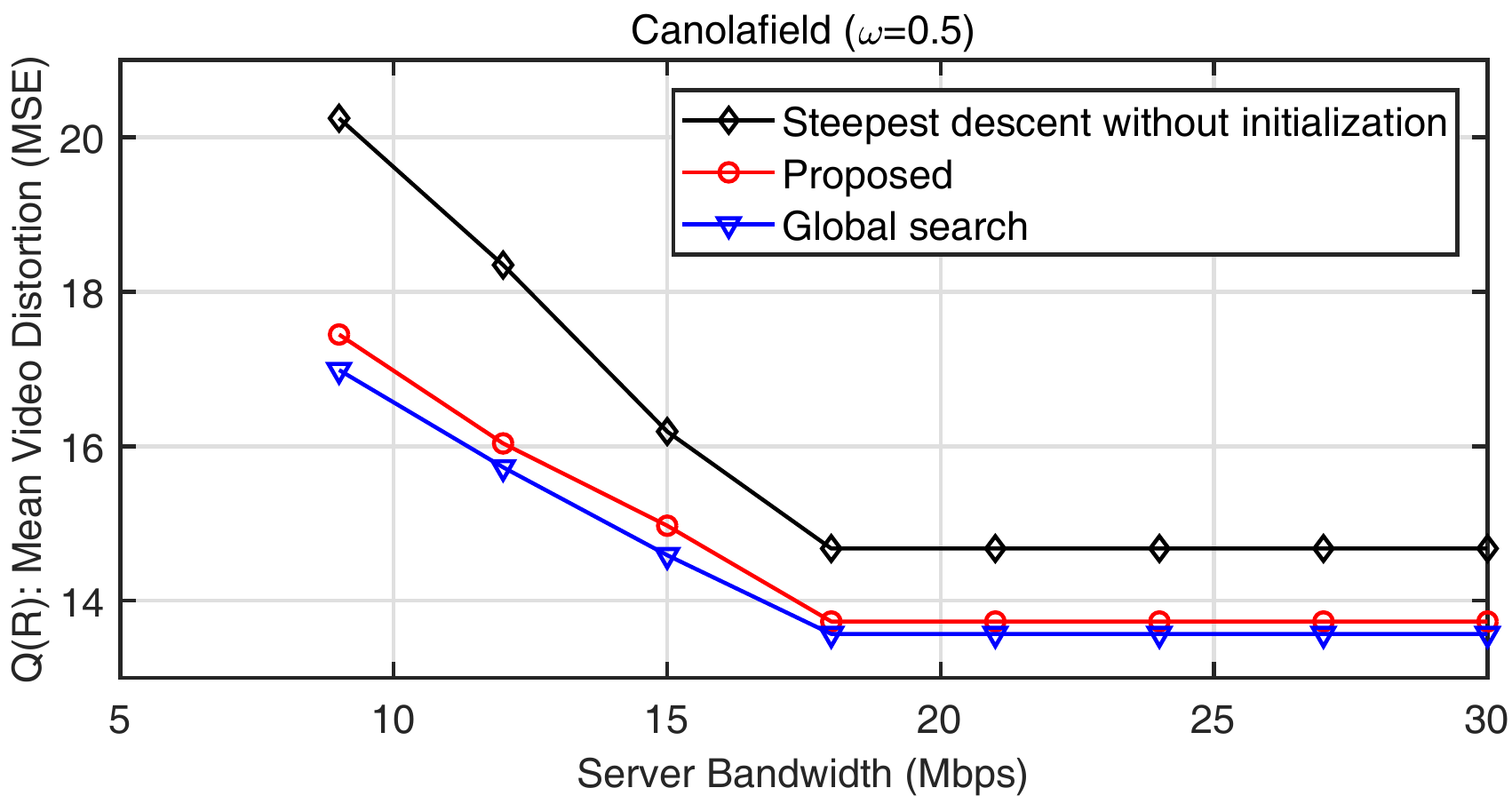} }
    \subfigure[]{ \label{fig_b}
    \includegraphics[width=3.05in]{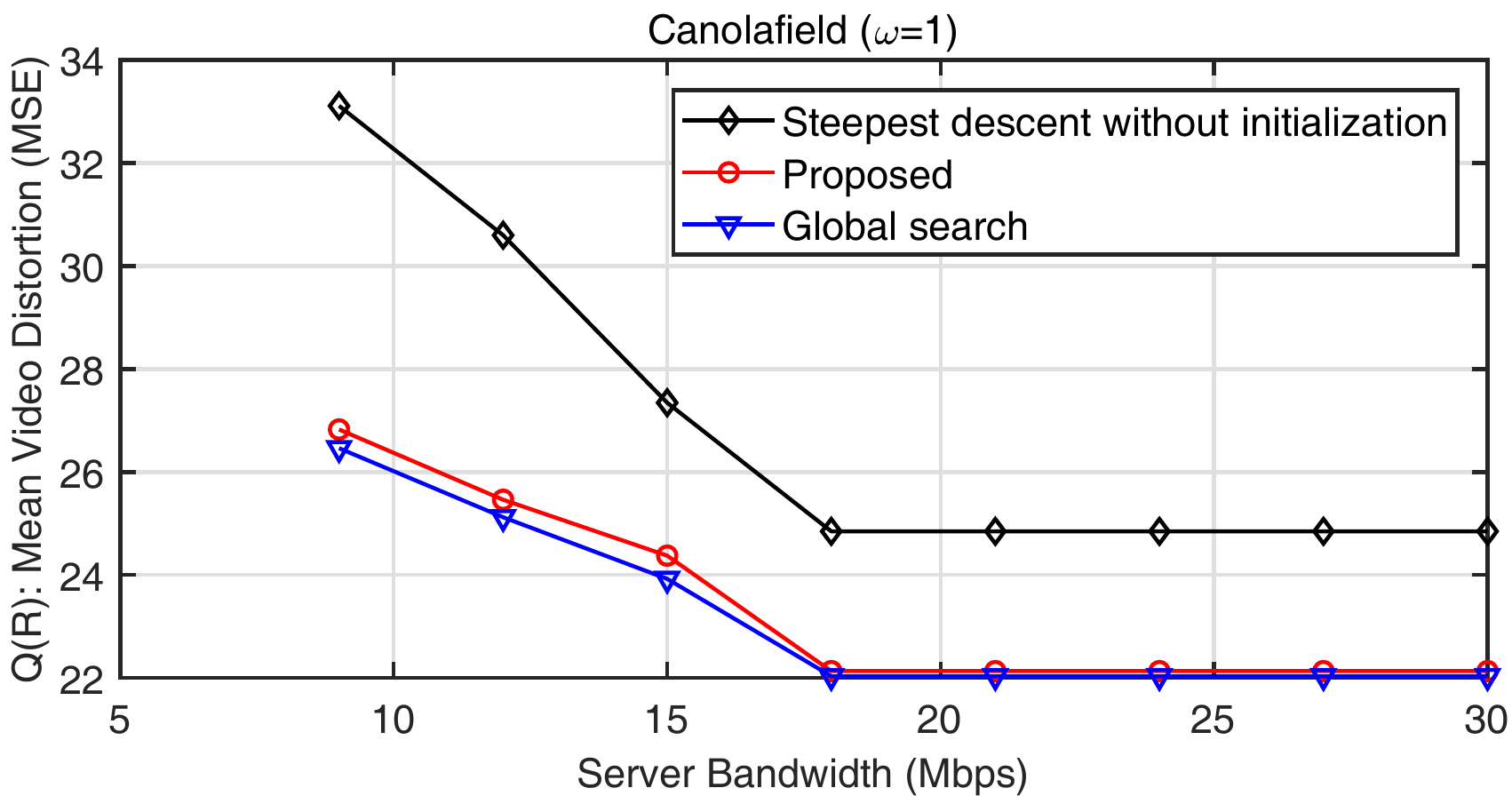} }
    \subfigure[]{ \label{fig_b}
    \includegraphics[width=3.05in]{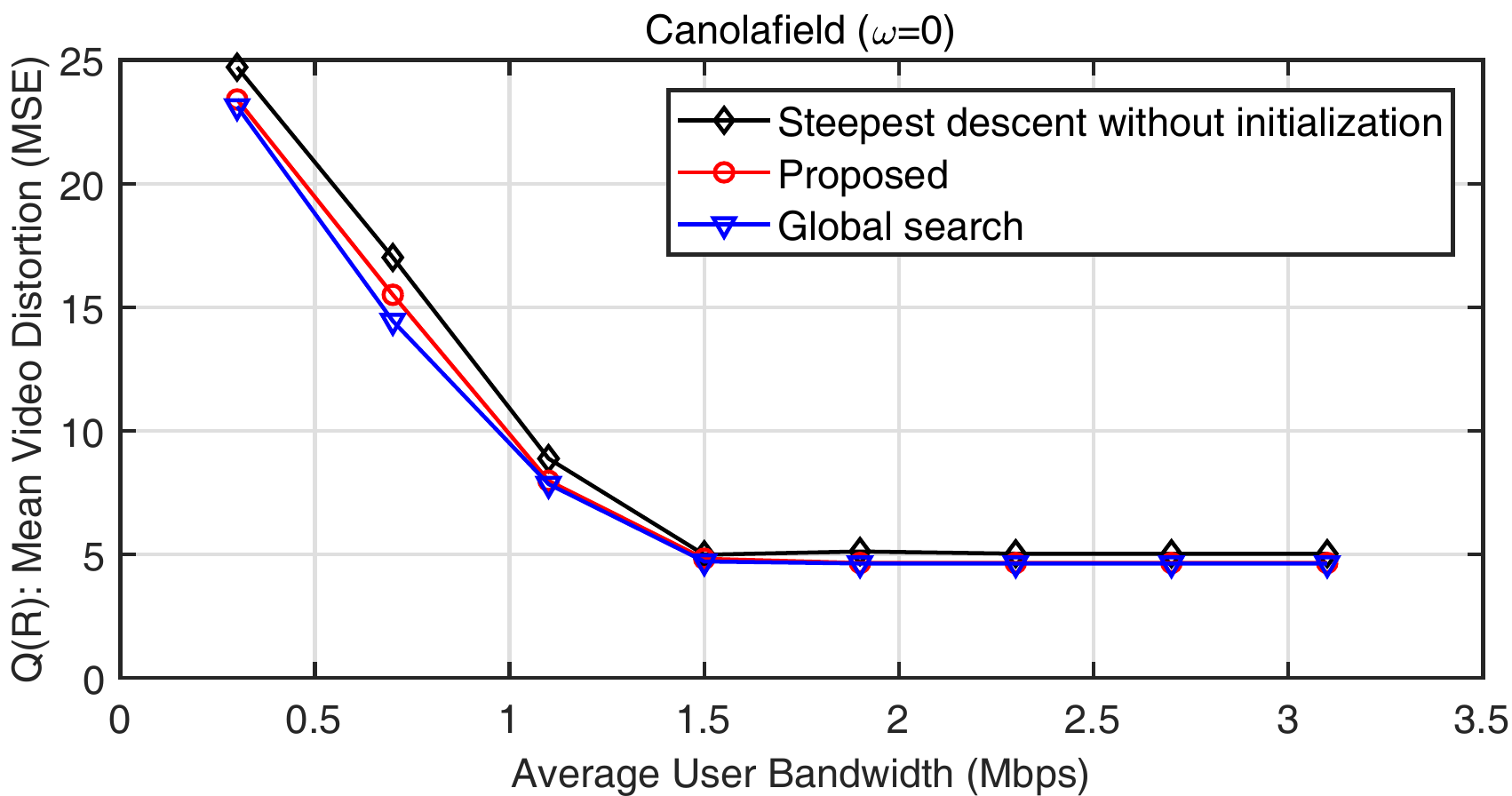} }
    \subfigure[]{ \label{fig_b}
    \includegraphics[width=3.05in]{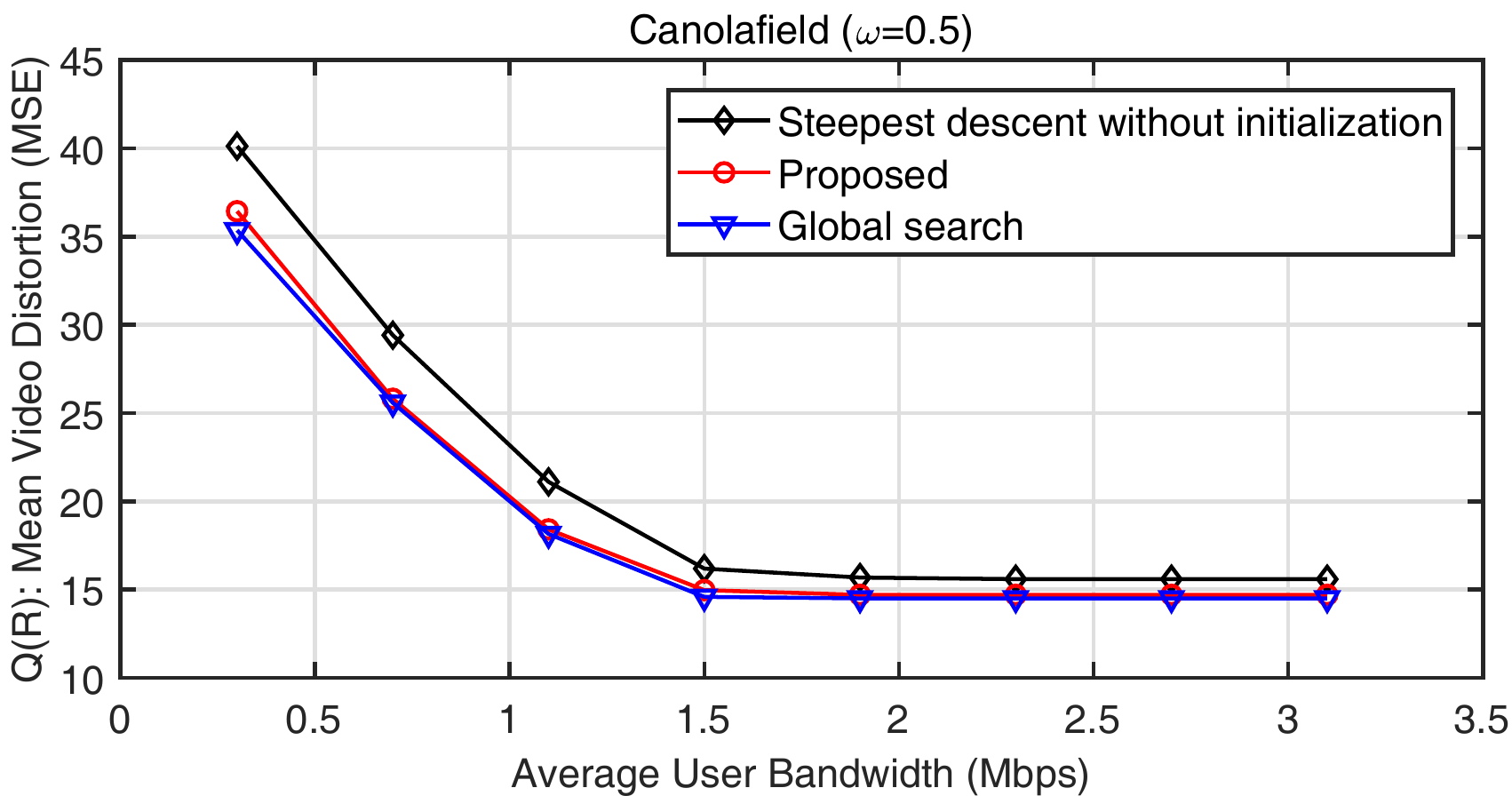} }
    \subfigure[]{ \label{fig_b}
    \includegraphics[width=3.05in]{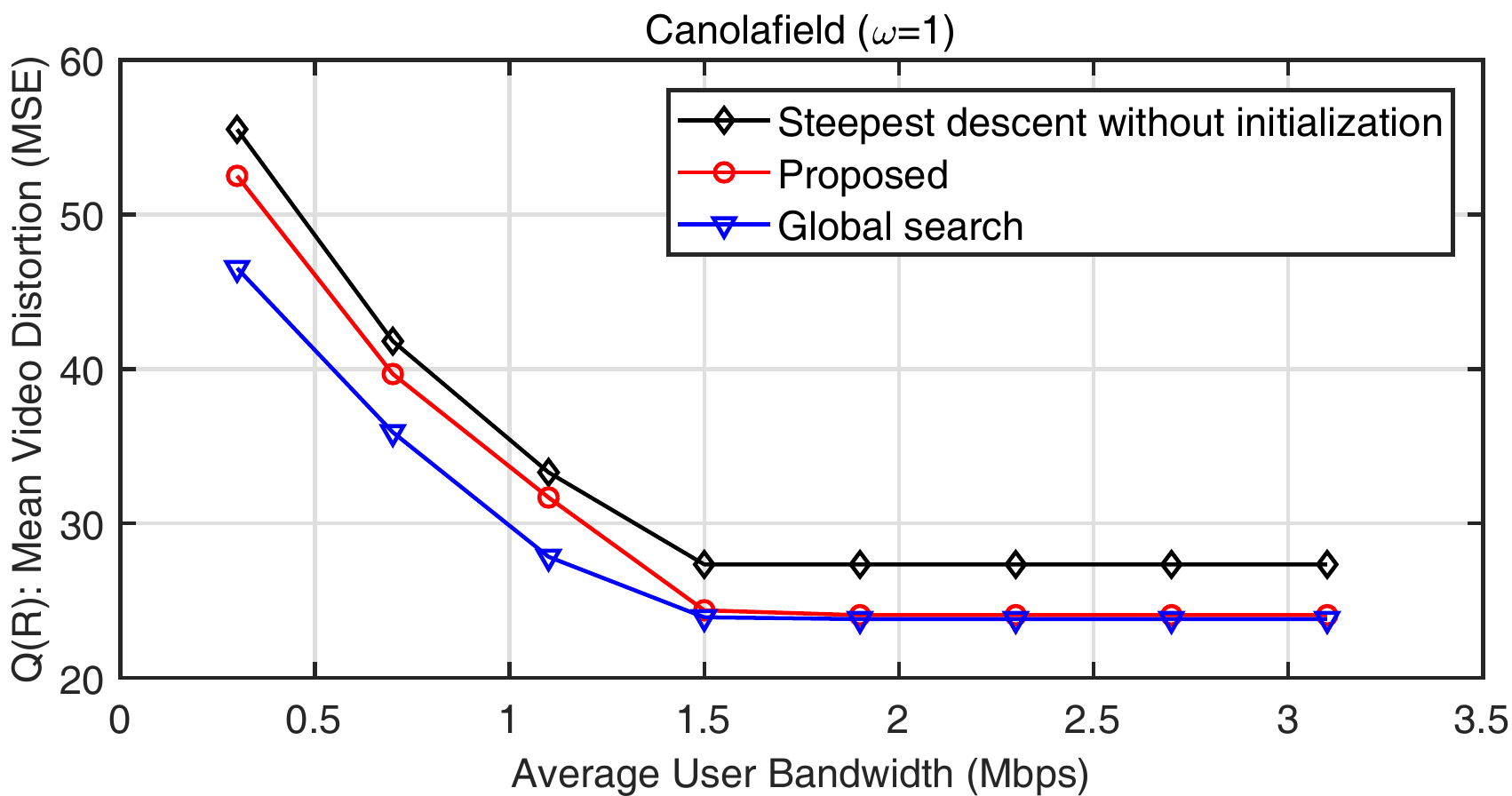} }
\caption{Comparison with the steepest descent solution without initialization, and the globally optimal solution achieved by global search.}
\label{fig:global}\vspace*{-0.7cm}
\end{figure*}

In Fig. \ref{fig:global}, we further compare the solution of problem \textbf{P1} in Eq. (\ref{eq:P1}) achieved by the global search approach, the steepest descent algorithm with initialization based on the continuous relaxation of problem \textbf{P1} as shown in Algorithm 1, and the steepest descent algorithm without such an initialization. In terms of the objective function value $Q(\mathbf{R})$, the global search approach can find the optimal solution of problem \textbf{P1} and thus returns the minimum value of $Q(\mathbf{R})$ for given transmission capacities of the server and users. For the steepest descent algorithm without initialization, the starting point for the search will be set as the lowest rate representation for each tile, i.e., $\mathbf{R}=\{R_1|\forall k, m,n\}$. This might cause the steepest descent search to drop into a local optimum, resulting in a solution much deviated from the optimal solution. In contrast, the proposed algorithm in Algorithm 1 utilizes the optimal solution of the continuous relaxation of problem \textbf{P1} to initialize the starting point for the steepest descent search, and obtains a solution much closer to the optimal solution. It can be seen in Fig. \ref{fig:global} that at some points, the proposed algorithm can even reach the optimal solution.

\section{Conclusion}
\label{sec:conslusion}

In this paper, we have investigated the server-side tile rate adaptation problem for multiple users competing for the server's transmission capacity. Based on the CNN-based viewpoint prediction, the mapping from the spherical viewport to its corresponding planar projection, and the corresponding visibility probability derivation of each tile for each user, it was then formulated as a non-linear discrete optimization problem to minimize the overall received video distortion of all users and the quality difference between the viewport and marginal tiles of each user, subject to the transmission capacity constraints of the server and users, and the specific viewport requirements of users. To solve this discrete optimization problem, we developed a steepest descent algorithm with the feasible starting point determined by the optimal solution of its continuous relaxation. Extensive simulations have been done under different system settings, empirically demonstrating the near-optimal performance of the proposed algorithm.



\end{document}